\documentclass[10pt,fleqn]{article}

\usepackage{amsmath,amssymb}
\usepackage{graphicx}
\usepackage[top=0.75in, bottom=0.75in, left=0.75in, right=0.75in, dvips]{geometry}

\begin{document}

\title{{\sf Stability of Subsequent-to-Leading-Logarithm Corrections to the Effective Potential for Radiative Electroweak Symmetry Breaking}}

\author{
F.A.\ Chishtie\thanks{Department of Applied Mathematics, The University of Western Ontario, London, Ontario  N6A 5B7, Canada},
V.\ Elias\thanks{Department of Applied Mathematics, The University of Western Ontario, London, Ontario  N6A 5B7, Canada},
R.B.\ Mann\thanks{
Department of Physics, University of Waterloo,
Waterloo, Ontario  N2L 3G1,  Canada}, 
D.G.C.\ McKeon\thanks{Department of Applied Mathematics, The University of Western Ontario, London, Ontario  N6A 5B7, Canada},
T.G.\ Steele\thanks{Department of Physics and Engineering Physics, University of
Saskatchewan, Saskatoon, SK, S7N 5E2, Canada}
}

\maketitle

\begin{abstract}
We demonstrate the stability under subsequent-to-leading logarithm corrections of the quartic scalar-field coupling constant $\lambda$ and the running Higgs boson mass obtained from the (initially massless) effective potential for radiatively broken electroweak symmetry in the single-Higgs-Doublet Standard Model. Such subsequent-to-leading logarithm contributions are systematically extracted from the renormalization group equation considered beyond one-loop order. 
We show $\lambda$ to be 
the dominant coupling constant of the effective potential for the radiatively broken case of electroweak symmetry.
We demonstrate the stability of $\lambda$ and the running Higgs boson mass through five orders of successively subleading logarithmic corrections to the scalar-field-theory projection of the effective potential for which all coupling constants except the dominant coupling constant $\lambda$ are disregarded. 
We present a full next-to-leading logarithm potential in the three dominant Standard Model coupling constants ($t$-quark-Yukawa, $\alpha_s$, and $\lambda$) from these coupling constants' contribution to two loop $\beta$- and $\gamma$-functions. Finally, we demonstrate the manifest order-by-order stability of the physical Higgs boson mass in the 220--231 GeV range.  In particular, we obtain a $231$ GeV physical Higgs boson mass inclusive of the $t$-quark-Yukawa and $\alpha_s$ coupling constants to 
next-to-leading logarithm order, and inclusive of the smaller $SU(2)\times U(1)$ gauge coupling constants to leading logarithm order.
\end{abstract}

\section{Introduction}
\renewcommand{\theequation}{1.\arabic{equation}}
\setcounter{equation}{0}
The motivation for radiative electroweak symmetry breaking, as first considered by Coleman and Weinberg \cite{1}, draws its roots all the way back to the first formulation of the hierarchy problem \cite{2} for embedding an $SU(2) \times U(1)$ electroweak gauge theory within a large grand unified theory.  As Sher points out in his review of radiative symmetry breaking \cite{3}, any scalar field mass term within the $SU(2) \times U(1)$ Lagrangian would necessarily be expected to have a magnitude sensitive to GUT-level mass scales via higher order processes involving the embedding theory.  However, indirect empirical bounds \cite{4} do not accommodate a Higgs boson mass appreciably larger than the electroweak vacuum expectation value $v = 246$ GeV.  Such a scalar field mass in conventional spontaneous symmetry breaking could occur only if the scalar-field mass term in the electroweak Lagrangian were exceedingly finely tuned to cancel off GUT-level scales from successive perturbative corrections arising within the embedding theory.  A more natural approach would be to assume that the embedding theory includes some symmetry ({\it e.g.} conformal invariance) that serves to protect the scalar-field mass term from such GUT-scale corrections.

Radiative symmetry breaking assumes this protective symmetry is \emph{exact}---that no scalar-field mass term appears in the electroweak Lagrangian. In the absence of such a mass term, Coleman and Weinberg \cite{1} found the one loop electroweak effective potential to be of the form
\begin{equation}
V_{eff}^{(1L)} = \phi^4 \left[ \frac{\lambda}{4} + \frac{9g_2^4 + 6g^{\prime^2} g_2^2 + 3g^{\prime^4} + 192 \lambda^2 - 48 g_Y^4}{1024 \pi^2} \log \left( \frac{\phi^2}{\mu^2} \right) + K\right],
\label{eq1.1}
\end{equation}
where $\lambda$ is the quartic scalar-field interaction coupling constant, $g_2$ and $g^\prime$ are the $SU(2) \times U(1)$ gauge interaction coupling constants, $g_Y$ is the dominant electroweak Yukawa coupling constant, $\mu$ is an arbitrary renormalization scale necessarily occurring as a by-product of the removal of infinities from $1$ loop graphs, and $K$ is the finite coefficient of $\phi^4$ after such ${\cal{O}}(\phi^4)$ infinities have been subtracted.

At the time of Coleman and Weinberg's paper, there was no evidence for a massive $t$-quark;  the magnitude of $g_Y$ could be reasonably assumed to be comparable to the $b$-quark Yukawa coupling constant $g_b^2 = 2(m_b / v)^2 \cong 8 \cdot10^{-4}$.  Any such comparable $g_Y^2$ would be ignorable relative to the corresponding electroweak gauge coupling constants $g_2^2 = e^2 / \sin^2 \theta_w \cong 0.44$, $g^{\prime^2} = e^2 / \cos^2 \theta_w \cong 0.13$.  Thus the gauge coupling constant terms were anticipated to dominate the opposite-sign Yukawa coupling constant term in the coefficient of the logarithm in (\ref{eq1.1}).

Coleman and Weinberg chose to determine the finite counterterm $K$ by insisting that $\mu$ be the scale at which $d^4 V_{eff}^{1L} / d\phi^4$ coincides with $d^4 V_{tree} / d\phi^4 = 6\lambda$. Ignoring $g_Y$ they found that 
\begin{equation}
K = -\frac{25}{6} \left[ \frac{9 g_2^4 + 6 g^{\prime^2} g_2^2 + 3g^{\prime^4} + 192 \lambda^2}{1024 \pi^2} \right] .
\label{eq1.2}
\end{equation}
Since $d^4V_{eff}^{(1L)} / d\phi^4$ is infinite at $\phi = 0$, Coleman and Weinberg necessarily had to choose a non-zero value of $\phi$ to serve as the scale of $V_{eff}^{(1L)}$.  
By identifying this scale with the electroweak vacuum expectation value $v$, {\it i.e.} by  identifying the arbitrary parameter $\mu$ in Eq.\ (\ref{eq1.1}) with the electroweak minimum $v$, one finds from the requirement $d V_{eff}^{(1L)} / d\phi|_v = 0$ that
\begin{equation}
\lambda = \frac{11}{256 \pi^2} \left[ 3g_2^4 + 2g^{\prime^2} g_2^2 + g^{\prime^4} + 64 \lambda^2 \right],
\label{eq1.3}
\end{equation}
where all coupling constants in (\ref{eq1.3}) are evaluated at the vev scale [{\it i.e.,} $\lambda = \lambda(v)$].  Although (\ref{eq1.3}) is formally a quadratic equation in $\lambda$, Coleman and Weinberg noted the existence of a small $\lambda$ solution 
$\left[ \lambda \cong \frac{11}{ 256 \pi^2} \left[ 3g_2^4 + 2g^{\prime^2} g_2^2 + g^{\prime^4} \right]  \right]$ for which radiative symmetry breaking would be perturbatively viable.

This approach, however, fails upon incorporation of the physical $t$-quark's Yukawa couplant:  $g_Y^2 = g_t^2 = 2 (m_t / v)^2 \cong 1.0$ in Eq.\  (\ref{eq1.1}).  Then the opposite-sign Yukawa coupling-constant term is seen {\it to dominate} the gauge coupling-constant terms, in which case equations (\ref{eq1.2}) and (\ref{eq1.3}) get replaced by relations
\begin{gather}
K \cong - \frac{25}{128 \pi^2} \left[ 4 \lambda^2 - g_t^4 \right],
\label{eq1.4}
\\
\lambda \cong  \frac{11}{16 \pi^2} \left[ 4 \lambda^2 - g_t^4 \right].
\label{eq1.5}
\end{gather}
As noted in ref.\ \cite{5}, there is no solution to (\ref{eq1.5}) for $\lambda$ sufficiently small to be perturbatively viable.  However, this does {\it not} mean that radiative electroweak symmetry breaking is impossible for empirical electroweak coupling constants.  Rather, it means that the one-loop effective potential (\ref{eq1.1}) is an inappropriate choice for radiative electroweak symmetry breaking;  leading-logarithm two-loop electroweak potential terms are comparable or larger than one-loop terms \cite{3}.  In refs.\ \cite{5} and \cite{6} it is argued that a potential based upon the {\it summation} of its leading-logarithm contributions leads not only to a Higgs boson mass within indirect empirical bounds \cite{6}, but also to a substantially reduced value for $\lambda$ that may be sufficiently small for perturbative stability of the physics extracted from the effective potential under its subsequent-to-leading-logarithm corrections.

In the present paper we address this issue by showing reasonable stability of the predictions of 
refs.\ \cite{5} and \cite{6} under subsequent-to-leading-logarithm corrections obtained via known renormalization-group functions for the electroweak couplings.  On a more general level, the present paper demonstrates how to formulate radiative symmetry breaking for effective potentials subject to a large destabilizing Yukawa couplant.  This is of value even if our specific choice of electroweak potential, which is considered in the present paper to arise from a single Higgs doublet, is incorrect.  We have seen how a large Yukawa couplant necessarily eliminates any small $\lambda$ ({\it i.e.} $\lambda \sim g_2^4$) solution.  The formulation of predictive results from radiative symmetry breaking when $\lambda$ is {\it not} fortuitously small, as in ref.\ \cite{1}, but rather the dominant couplant in the effective potential, is important both for electroweak symmetry breaking and for cosmological applications \cite{7}.

To develop these ideas further, we first review in Section 2 the leading logarithm results of 
refs.\ \cite{5} and \cite{6}.  These results, incorporating the largest Standard Model couplants
\begin{gather}
x = g_t^2 (v) / 4 \pi^2, \; \; y = \lambda (v) / 4\pi^2, \; \; z = \alpha_s (v) / \pi,
\\
r = g_2^2 (v) / 4\pi^2, \; {\rm and}\; \;  s = \left(g^\prime (v) \right)^2 / 4\pi^2,
\end{gather}
are indicative of a Higgs boson mass of 218 GeV, in conjunction with the dominance of the scalar self-interaction couplant $y$ over all other Standard Model couplants.

In Section 3, we consider the scalar field theory projection (SFTP) of the effective potential, the approximation in which all Standard Model couplants except $y$ are ignored.  Such a theory has renormalization group functions $\beta_y (y)$ and $\gamma(y)$ that are known to 5-loop order.  We are therefore able to construct successive approximations to the full SFTP involving the summation of leading and four successively subleading logarithms in the full effective potential series.

Since the SFTP is not scale-free (by virtue of gauge-sector interactions, its vacuum expectation value $v$ is constrained to satisfy $M_W = g_2 v/2$, or alternatively, $v = G_F^{-1/2} 2^{-1/4}$), we are able to calculate the running Higgs boson mass $\left[ V^{\prime\prime} (v) \right]^{1/2}$ for each successive order.  We find surprisingly that the leading logarithm SFTP reproduces quite closely the full leading logarithm results of Section 2, despite the deliberate omission of other Standard Model couplants $\{x, z, r, s\}$ from the SFTP potential.

In Section 4, we develop an explicit set of successive subleading-logarithm approximations to the full SFTP potential.  We then demonstrate for this potential that the running Higgs boson mass $\left[V^{\prime\prime}(v)\right]^{1/2} (\cong 226$ GeV) and the scalar field self-interaction couplant $y(v) (\cong 0.054$) are remarkably stable as the order of these approximations increases.

In Section 5, we demonstrate how known two-loop renormalization group functions involving the dominant Standard Model couplants $\{x, y, z\}$ can be utilized to obtain leading and next-to-leading logarithm corrections to the SFTP resulting from ``turning on'' the $t$-quark Yukawa interaction.  Thus, we are able to include the next-to-leading logarithm corrections to the leading-logarithm potential of 
refs.\ \cite{5} and \cite{6}.

In Section 6 we find such corrections alter the running Higgs boson mass obtained from the SFTP potential by only a few GeV, depending upon the order of the SFTP employed. In particular, we find the fully $NLL$ extension of $LL$ effective potential expressed in terms of dominant $SM$ couplants $x,y,z$ \cite{5,6} leads to a running Higgs boson mass of $228$ GeV and a corresponding couplant $y(v) = 0.0531$.

Finally, in Section 7, we discuss further aspects of subleading logarithm corrections to the effective potential series.  We first consider augmentation of the results of Section 6
by the leading-logarithm contributions of the relatively small electroweak gauge couplants $r$ and $s$ to the effective potential.  Upon incorporation of these contributions, the running Higgs boson mass is found to be quite stable at $~224$ GeV over successive orders of SFTP scaffolding. The couplant $y (v) \cong 0.054 $ is similarly shown to exhibit stability over four SFTP orders. The fully $NLL$ results in $\{x,y,z\}$ described in the previous paragraph are altered slightly upon incorporation of $LL$ corrections in $\{r,s\}$ to a Higgs boson mass of $231$ GeV, with $y(v) = 0.054$.
We also show that the physical Higgs boson mass differs by less than $0.3$ GeV from the mass obtained from the effective potential taken to $NLL$ order in contributions from $g_t$ and $\alpha_s$, and to $LL$ order in contributions from the $SU(2) \times U(1)$ gauge coupling constants. We discuss the phenomenological differences anticipated between radiatively and conventionally broken electroweak symmetry. We find surprising lowest-order agreement between both approaches for a number of processes [$WW \rightarrow (ZZ, WW); \; \; H \rightarrow (ZZ, WW)$], though we do find an enhancement of  two-Higgs scattering processes like $WW\to HH$ for the radiative case over the conventional symmetry breaking case.  We conclude with a summary of the methodology employed toward obtaining order-by-order stability in the physics extracted from radiative electroweak symmetry breaking. 



\renewcommand{\theequation}{2.\arabic{equation}}
\setcounter{equation}{0}

\section{Review of Leading Logarithm Results}
In refs.\ \cite{5,6}, the summation of leading-logarithm contributions to the effective potential of the Standard Model is expressed in terms of its dominant three couplants
\begin{equation}
x = g_t^2 (v) / 4\pi^2 = 0.0253, \; y = \lambda (v) / 4\pi^2, \; z = \alpha_s (v)/ \pi = 0.0329,
\label{eq2.1}
\end{equation}
where $v = \langle \phi \rangle$, the vacuum expectation value.
The leading logarithm contribution is of the series form 
\begin{gather}
  \pi^2 \phi^4 S_{LL} = \pi^2 \phi^4 \left\{ \sum_{n=0}^\infty x^n \sum_{k=0}^\infty y^k \sum_{\ell = 0}^\infty z^\ell 
C_{n,k,\ell} L^{n+k+\ell-1} \right\}, 
\label{eq2.2}
\\ 
 L \equiv \log (\phi^2 / v^2 ), \; \;  C_{0,0,0} = C_{1,0,0} = C_{0,0,1} = 0, \; C_{0,1,0} = 1.
\nonumber
\end{gather}
This effective potential is shown in ref.\ \cite{6} to include explicitly the one loop contributions $C_{0,2,0} = 3$, $C_{2,0,0} = - 3/4$ [all other $C_{i,,j,k}$ with $i+j+k=2$ are equal to zero] derived from Feynman graphs in ref.\ \cite{1}.  All remaining coefficients $C_{n,k,\ell}$ in (\ref{eq2.2}) are determined by the renormalization group (RG) equation, in which all RG functions are evaluated to one-loop order:
\begin{equation}
\left[ (-2-2\gamma) \frac{\partial}{\partial L} + \beta_x \frac{\partial}{\partial x} + \beta_y \frac{\partial}{\partial y} + \beta_z \frac{\partial}{\partial z} - 4\gamma \right] S = 0.
\label{eq2.3}
\end{equation}

Only one-loop RG functions enter, because $S_{LL}$ is determined in full by those contributions to the differential operator on the left hand side of (\ref{eq2.3}) that either lower the degree of $L$ by one or raise the aggregate power of couplants by one \cite{5,6}.  The leading logarithm effective potential may be expressed as a power series in the logarithm $L$:\footnote{We also note that the RG equation (\protect\ref{eq2.3}) can be applied directly to the expression  Eq.\ (\protect\ref{eq2.4}) as in \cite{17}.}
\begin{equation}
V_{eff}  =  \pi^2 \phi^4 \left( S_{LL} + K \right)
= \pi^2 \phi^4 \left( A+BL+CL^2+DL^3+EL^4 + \ldots \right),
\label{eq2.4}
\end{equation}
where the constant $K$ includes all finite $\phi^4$ counterterms remaining after divergent contributions from $\phi^4$ graphs degree-2 and higher in couplant powers are removed.  Thus the purely $\phi^4$ coefficient $A$ is equal to $y+K$, and coefficients $\{B, C, D, E \}$ are explicitly obtained in 
refs.\ \cite{5,6} via Eq.\  (\ref{eq2.3}) as degree $\{2,3,4,5 \}$ polynomials in the couplants $x,y,z$.  The unknown couplant $y(v)$ and
finite counterterm $\pi^2 K \phi^4$ in $V_{eff}$ are
numerically determined by the simultaneous application of Coleman and Weinberg's renormalization conditions \cite{1}, which impose a minimum at $\phi = v$ and which define the quartic scalar-field interaction couplant $\lambda$ at $\phi = v$: \footnote{For the {\it one-loop} potential of Coleman and Weinberg (in which $C=D=E=0$), these conditions imply that $K = -25B/6$, $y = 11B/3$, consistent with Eqs.\  (\ref{eq1.2}) and (\ref{eq1.3}).} 
\begin{gather}
V_{eff}^\prime (v) = 0 \Rightarrow K = - B/2 - y,
\label{eq2.5}
\\
V_{eff}^{(4)} (v) = \frac{d^4}{d\phi^4} \left( \lambda \phi^4 \right) = 24\pi^2 y \Rightarrow y = \frac{11}{3} B + \frac{35}{3} C + 20D + 16E.
\label{eq2.6}
\end{gather}
In terms of the dominant three couplants (\ref{eq2.1}), one finds from Eqs.\  (\ref{eq2.5}) and (\ref{eq2.6}) that 
{\allowdisplaybreaks
\begin{gather}
K = -y + 3x^2 / 8 - 3y^2/2
\label{eq2.7}
\\
\begin{split}
y  = & \left[ 11y^2 - \frac{11}{4} x^2 \right] + \left[ 105 y^3 + \frac{105}{4} \left( xy^2 - x^2 y\right) + \frac{35}{2} x^2 z - \frac{105}{32} x^3 \right] 
\\
& +  \left[ 540y^4 + 270xy^3 - 30xy^2z + 60x^2yz - \frac{1125}{8} x^2 y^2
 -   \frac{115}{2} x^2 z^2 + \frac{75}{4} x^3 z - \frac{225}{4}x^3 y + \frac{495}{64} x^4 \right]
\\
& +  \left[ 1296 y^5 + 972 xy^4 - 144 x y^3 z + \frac{45}{2} xy^2z^2
 -  69 x^2 yz^2 - 270 x^2 y^3 + \frac{531}{4} x^2 y^2 z + \frac{345}{4} x^2 z^3\right.
\\
&\qquad \left.-  \frac{603}{16} x^3 z^2 + \frac{207}{2} x^3 yz - \frac{8343}{32} x^3 y^2
 -   \frac{459}{32} x^4 z + \frac{135}{32} x^4 y + \frac{837}{64} x^5 \right],
\end{split}
\label{eq2.8}
\end{gather}
}where $x = x(v) = 0.0253$ and $z = z(v) = 0.0329$ in the above two equations.  Thus equation (\ref{eq2.8}) is a fifth order polynomial equation for $y(\equiv y(v))$, with solutions $y = \{0.0538, \; -0.278, \; -0.00143 \}$.  Since $y$ must be a positive-definite couplant, only the first of these solutions is viable ($y(v)=0.053829$), in which case the finite counterterm $K$ is found from Eq.\  (\ref{eq2.7}) to be $-0.057935$.  The running Higgs boson mass \cite{8}, which is defined from the second derivative of the effective potential at its minimum $v$,
\begin{equation}
m_H^2 = V_{eff}^{\prime\prime} (v) = 8\pi^2 v^2 (B+C),
\label{eq2.9}
\end{equation}
is found from the physical values (\ref{eq2.1}) for $x(v)$ and $z(v)$ and the numerical solution to Eq.\  (\ref{eq2.8}) $y(v) = 0.0538$ to be $m_H = 216$ GeV.

When one augments the dominant couplants (\ref{eq2.1}) with the next-largest couplants in the Standard Model, the electroweak gauge couplants
\begin{equation}
r = \left[ g_2 (v) \right]^2 / 4\pi^2 = 0.0109, \; s = \left[ g^\prime (v) \right]^2 / 4\pi^2 = 0.00324,
\label{eq2.10}
\end{equation}
one finds from the additional $r$- and $s$-dependent contributions to $\{B, C, D, E \}$, as listed in the erratum to ref.\ \cite{6}, that the numerical solution to Eq.\ (\ref{eq2.6}) for $y(v)$ is altered slightly from $0.05383$ to $0.05448$.  Correspondingly, we find from Eq.\ (\ref{eq2.5}) that the finite counterterm $K$ is also altered from $-0.0579$ to
\begin{equation}
K = -y + 3x^2 / 8 - 3y^2 / 2 - 3s^2 / 128 - 9r^2 / 128 - 3rs / 64 = -0.058703.
\label{eq2.11}
\end{equation}
Most importantly, however, the running Higgs boson mass (\ref{eq2.9}) is elevated only minimally ($m_H = 218$ GeV) from its $216$ GeV value when electroweak gauge couplants are omitted.  Thus, the leading logarithm effective potential (\ref{eq2.2}) results appear to be stable under the incorporation of the leading additional subdominant electroweak couplants, {\it i.e.,} the electroweak gauge couplants themselves.

The question that remains, however, is whether the value determined for $y(v)$ ($=0.0545$) is sufficiently small for these leading-logarithm predictions to be reasonably stable under subsequent-to-leading-logarithm corrections to the scalar field potential.  The couplant $y(v)$ is clearly still the dominant electroweak couplant in the radiative symmetry-breaking scenario, as is evident by comparison to couplants $x$ and $z$ [Eq.\  (\ref{eq2.1})].  Indeed, such a value for $y$ ($=\lambda / 4\pi^2$) would correspond to having a Higgs boson mass of $510$ GeV in a conventional symmetry breaking scenario.  Nevertheless, the two-loop terms in known electroweak $RG$ functions are still seen to be substantially smaller than the one-loop terms when $y(v) = 0.0545$ \cite{5}, suggesting that corrections to $y$ and $m_H$ from subsequent-to-leading logarithms may be controllable in radiatively broken electroweak symmetry.



\renewcommand{\theequation}{3.\arabic{equation}}
\setcounter{equation}{0}

\section{The Scalar Field Theory Projection of $V_{eff}$}
To address the stability of leading logarithm results for electroweak symmetry breaking, it is useful to first consider the scalar-field-theory projection (SFTP) of the electroweak effective potential.  This projection corresponds to the potential one obtains by omitting all standard model (SM) couplants except for the dominant scalar-field self-interaction couplant $y$.  Such an approach is analogous to the usual SM procedure for processes (such as $R(s)$) involving both strong and electroweak perturbative corrections:  one first evaluates QCD corrections in isolation, since $\alpha_s$ is the dominant coupling constant, prior to introducing SM corrections from subdominant electroweak gauge coupling constants.

The summation-of-leading-logarithms SFTP for radiative electroweak symmetry breaking is known to {\it all} orders in the couplant $y$, and is given in closed form as the $x = 0$ limit of Eq.\  (6.1) in ref.\ \cite{6}:
\begin{equation}
V_{SFTP}^{LL} = \pi^2 \phi^4 \left[ \frac{y}{1-3y L} + K  \right] .
\label{eq3.1}
\end{equation}
The constant $K$ is the residual coefficient of $\phi^4$ after infinities from multiloop $\phi^4$ graphs have been subtracted.  Thus $K$ is inclusive of all finite counterterms degree-2 and higher in $y$.  Curiously, one finds by applying the condition (\ref{eq2.6}) to the series expansion of (\ref{eq3.1}),
\begin{equation}
y = 11y^2 + 105y^3 + 540y^4 + 1296y^5,
\label{eq3.2}
\end{equation}
that a solution $y = 0.054135$ exists quite close to the one quoted in the previous section ($y = 0.0545$) when nonzero physical values are included for the subdominant electroweak couplants $x$, $z$, $r$ and $s$.  Similarly, one finds from the condition (\ref{eq2.5}) that $K = -y - 3y^2 / 2 = - 0.05853$, in close agreement with the aggregate counterterm coefficient (\ref{eq2.11}) when the same subdominant electroweak couplants are included.  Although we are approximating all subdominant electroweak couplants to be zero in the SFTP potential, this potential is not scale-free;  a physical $vev$-scale $v = 2^{-1/4} G_F^{-1/2}$ still arises from the SM gauge sector.
We require the vacuum expectation value of the SFTP potential to be at $v = 246.2$ GeV, and then find from Eq.\ (\ref{eq2.9}) that $m_H = 221.2$ GeV [$B = 3y^2, \; C = 9y^3, \; y = 0.054135$], only a small departure from the $218$ GeV result \cite{6} obtained when the subdominant couplants $x, z, r, s$ are no longer omitted, but ``turned on'' from zero to their physical values (0.0253, 0.0329, 0.0109, 0.00324, respectively).  These results demonstrate that the SFTP approximation is a surprisingly good one for leading-logarithm radiative electroweak symmetry breaking--- that $y$ is truly the driving couplant for obtaining the leading-logarithm results summarized in the previous section.

To carry the SFTP approximation (in the absence of an explicit scalar-field mass term) past leading-logarithms, we first note that the SFTP {\it all-orders} potential takes the form of a perturbative field theoretic series ($y = \lambda/4\pi^2, {\cal{L}} = \log (\phi^2 / \mu^2)$)
\begin{equation}
V_{SFTP} = \pi^2 \phi^4 S_{SFTP}, \; \; S_{SFTP} = y + \sum_{n=1}^\infty \sum_{m=0}^n T_{n,m} y^{n+1} {\cal{L}}^m.
\label{eq3.3}
\end{equation}
Leading-logarithm contributions to this series involve coefficients $T_{n,n} = 3^n$, as is evident from the leading-logarithm potential (\ref{eq3.1}).  Thus, the potential (\ref{eq3.1}) includes only $m=n$ terms of $V_{SFTP}$ (\ref{eq3.3}), as well as $n \geq 1 \; \; , m=0, \;  \phi^4$-counterterms absorbed in $K$.
If ${\cal{L}} \rightarrow L$ ({\it i.e.} $\mu \rightarrow v$), we see that the potential (\ref{eq3.1}) is the leading logarithm projection of the full SFTP potential (\ref{eq3.3}), provided that the finite $\phi^4$ counterterm coefficient $K$ in Eq.\  (\ref{eq3.1}) corresponds to the aggregate contribution of all purely $\phi^4$-terms in $V_{SFTP}$ that are degree-2 and higher in $y$:
\begin{equation}
K = \sum_{n=1}^\infty y^{n+1}T_{n,0} .
\label{eq3.4}
\end{equation}
In other words, the finite counterterm coefficient $K$ in Eq.\  (\ref{eq3.1}) is inclusive of all terms in the full SFTP potential that can contribute to it.

The invariance of $V_{SFTP}$ under changes in the renormalization scale $\mu$ implies that $S_{SFTP}$ satisfies the RGE
\begin{equation}
\left[ (-2-2\gamma) \frac{\partial}{\partial {\cal{L}}} + \beta_y \frac{\partial}{\partial y} - 4\gamma \right] \; S = 0,
\label{eq3.5}
\end{equation}
where $RG$ functions $\beta_y$ and $\gamma$ have been calculated to 5-loop order for global $O(N)$ symmetric scalar field theory \cite{9}.  The $SM$ $RG$ functions in the SFTP of the single Higgs effective potential are just the $N = 4$ case of this theory:
{\allowdisplaybreaks
\begin{gather}
\beta_y = 6y^2 - \frac{39}{2} y^3 + 187.855 y^4 - 2698.27 y^5 + 47974.7 y^6 + \ldots
\label{eq3.6}
\\
\gamma = \frac{3}{8} y^2 - \frac{9}{16} y^3 + \frac{585}{128} y^4 - 49.8345 y^5 + \ldots
\label{eq3.7}
\end{gather}
}
The series $S_{SFTP}$ in the full potential (\ref{eq3.3}) may be rewritten in terms of {\it sums} of leading ($S_0$) and successively subleading ($S_1, S_2, \ldots$) logarithms:
\begin{equation}
S_{SFTP} = y S_0 (y{\cal{L}}) + y^2 S_1 (y{\cal{L}}) +y^3 S_2 (y{\cal{L}}) 
+y^4 S_3 (y{\cal{L}}) + \ldots ; \; \; \; \; S_k (u) \equiv \sum_{n=k}^\infty T_{n,n-k} u^{n-k}.
\label{eq3.8}
\end{equation}

Given $u = y{\cal{L}}$, we employ the methods of ref.\ \cite{10} to obtain successive differential equations for $S_k (u)$, first by substituting Eq.\  (\ref{eq3.8}) into the $RG$ equation (\ref{eq3.5}), and then by organizing the $RG$ equation in powers of $y$:
\begin{gather}
{\cal{O}}(y^2): 2(1-3u) \frac{d S_0}{du} - 6S_0 = 0, \; \; S_0 (0) = T_{0,0} = 1;
\label{eq3.9}
\\
{\cal{O}}(y^3): 2(1-3u) \frac{d S_1}{du} - 12S_1 = -21 S_0 - \frac{39}{2} u \frac{d S_0}{du},  \; \; S_1 (0) = T_{1,0};
\label{eq3.10}
\\
{\cal{O}}(y^4): 2(1-3u) \frac{d S_2}{du} - 18S_2  =  190.105 S_0 - \frac{3}{4}  \frac{d S_0}{du} + 187.855 u \frac{d S_0}{du} - \frac{81}{2} S_1 - \frac{39}{2} u \frac{d S_1}{du},
\qquad S_2 (0) = T_{2,0};
\label{eq3.11}
\\
\begin{split}
{\cal{O}}(y^5):  2(1 - 3u) \frac{d S_3}{du} - 24 S_3 
 = & -2716.55 S_0 + \frac{9}{8} \frac{d S_0}{du} - 2698.27 u \frac{d S_0}{du}\\
 &+  377.959 S_1 - \frac{3}{4} \frac{d S_1}{du} + 187.855 u \frac{d S_1}{du}
  -  60 S_2 - \frac{39}{2} u \frac{d S_2}{du}, \; \; S_3 (0) = T_{3,0};
\end{split}
\label{eq3.12}
\\
\begin{split}
{\cal{O}}(y^6):  2(1-3u) \frac{d S_4}{du} - 30 S_4
 = & 48174.1 S_0 - \frac{585}{64} \frac{d S_0}{du} + 47974.7 u \frac{d S_0}{du}
 -  5414.81 S_1 + \frac{9}{8} \frac{d S_1}{du} - 2698.27 u \frac{d S_1}{du} 
 \\
& +  565.814 S_2 - \frac{3}{4} \frac{d S_2}{du} + 187.855 u \frac{d S_2}{du}
 -  \frac{159}{2} S_3 - \frac{39}{2} u \frac{d S_3}{du}, \; \; S_4 (0) = T_{4,0}.
\end{split}
\label{eq3.13}
\end{gather}
Equations for summations $S_k$ with $k > 4$ require the knowledge of 6-loop-and-higher terms in the $RG$ functions (\ref{eq3.6}) and (\ref{eq3.7}).

Initial conditions for all but $S_0$ are dependent on finite
coefficients $T_{k,0}\; (k > 0)$ of $\phi^4$ after infinities are subtracted.  Such coefficients will be determined via successive applications of Eq.\  (\ref{eq2.5}). The solution to Eq.\  (\ref{eq3.9}) is
\begin{equation}
S_0 (u) = \frac{1}{1-3u},
\label{eq3.14}
\end{equation}
recovering $(u = yL)$ the leading logarithm sum within the potential (\ref{eq3.1}).  Similarly one finds explicit solutions to Eqs.\  (\ref{eq3.10})--(\ref{eq3.13}) to be
{\allowdisplaybreaks
\begin{gather}
S_1 (u) = \frac{4 T_{1,0} - 3u + 13 \log (1 - 3u)}{4(1 - 3u)^2},
\label{eq3.15}
\\
\begin{split}
(1 - 3u)^3 S_2 (u)  = & T_{2,0} + \frac{9u^2}{4} + u \left( -\frac{3}{4}T_{1,0} +59.8023 \right)
 +  \frac{13}{16} (1 - 3u) \log (1 - 3u) 
\\
&+ \left( \frac{13}{2} T_{1,0} - \frac{195}{16} \right) \log (1 - 3u)
 +   \frac{169}{16} \left[\log (1 - 3u) \right]^2   .
\end{split}
\label{eq3.16}
\\
\begin{split}
(1 - 3u)^4 S_3 (u)=& T_{3,0} +\left(-\frac{3}{4}T_{2,0}-1047.88+120.917T_{1,0}  \right)u 
+\left(\frac{9}{4}T_{1,0}+1190.26  \right) u^2
+28.6798u^3
\\
& +\left[101.754+\frac{39}{4}T_{2,0}-\frac{351}{16} T_{1,0} 
+\left(401.512-\frac{39}{8}T_{1,0} \right) u
+\frac{117}{16}u^2
\right]\log{\left(1-3u\right)}
\\
&+\left(-\frac{11661}{128}-\frac{507}{64}u +\frac{507}{16}T_{1,0}\right)\left[\log{\left(1-3u\right)}\right]^2
+\frac{2197}{64}\left[\log{\left(1-3u\right)}\right]^3
\end{split}
\label{s3}
\\
\begin{split}
(1 - 3u)^5S_4 (u)=& T_{4,0}
+\left(19636.1-\frac{3}{4}T_{3,0}+182.032\,T_{2,0}
-2089.57\,T_{1,0} \right)u 
\\
&+\left(-44052.4+2409.76\,T_{1,0}+\frac{9}{4}T_{2,0} \right)u^2
+\left(28.6798\,T_{1,0}+42536.2 \right) u^3
+1085.32u^4
\\
&+\Biggl\{93.2095u^3+\left(\frac{117}{8}T_{1,0}+7806.13\right)u^2\Biggr.
\\
&\qquad\qquad
+\left(-8938.02+1199.66\,T_{1,0}-\frac{117}{16}T_{2,0}\right) u
\\
&\qquad\qquad+\Biggl.202.290\,T_{1,0}-1465.37+13 T_{3,0}-\frac{65}{2}T_{2,0}
\Biggr\}\log{\left(1-3u\right)}
\\
&+\Biggl\{
\frac{1521}{64}u^2+\left(-\frac{1521}{64}T_{1,0}+1991.04 \right)u\Biggr.
\\
&\qquad\qquad\Biggl.+1174.97-\frac{7943}{32}T_{1,0}+\frac{507}{8}T_{2,0}
\Biggr\}\left[\log{\left(1-3u\right)}\right]^2
\\
&+\left\{ -\frac{6591}{256}u-\frac{195533}{384}+\frac{2197}{16} T_{1,0}
\right\}\left[\log{\left(1-3u\right)}\right]^3
+\frac{28561}{256}\left[\log{\left(1-3u\right)}\right]^4
\end{split}
\label{s4}
\end{gather}
}
Obtaining $S_k (u)$ with $k > 4$ requires knowledge past 5-loop order terms in the $RG$ functions (\ref{eq3.6}) and (\ref{eq3.7}), as noted above.  We see, however, that if $\mu$ is chosen to equal the electroweak $vev$ scale $v = 246.2$ GeV, the all-orders SFTP potential (\ref{eq3.3}) may be expressed in the form [$L = \log \phi^2 / v^2$ as before; $y = y(v)$]
\begin{equation}
V_{SFTP} = \pi^2 \phi^4 y \left[ S_0 (yL) + y S_1 (yL) + y^2 S_2 (yL) + y^3 S_3 (yL) + 
y^4 S_4 (yL) +
\ldots \right]
\label{eq3.17}
\end{equation}
where $S_k (yL)$ is given by the final expression of Eq.\  (\ref{eq3.8}).



\section{Successive Approximations to the Full SFTP Potential}

\renewcommand{\theequation}{4.\arabic{equation}}
\setcounter{equation}{0}

The all-orders SFTP Potential, expressed as the double summation (\ref{eq3.3}) with ${\cal{L}} \rightarrow L \; (\mu \rightarrow v)$ may be approached by successive summations of subleading logarithms [$LL \equiv$ leading-$\log$, $N^k LL \equiv$ (next-to-)$^k$-leading $\log$] contributing to the complete perturbative series (\ref{eq3.16}):
\begin{gather}
V_{LL}  =  \pi^2 \phi^4 y \left[ \sum_{n=0}^\infty T_{n,n} (yL)^n + y T_{1,0} + y^2 T_{2,0} + y^3 T_{3,0} + y^4 T_{4,0} + \ldots\right]
 =  \pi^2 \phi^4 \left[ y S_0 (yL) + K \right]
 \label{eq4.1}
 \\
\begin{split}
V_{NLL} & =  \pi^2 \phi^4 y \left[ \sum_{n=0}^\infty T_{n,n} (yL)^n + y \sum_{n=1}^\infty T_{n,n-1} (yL)^{n-1}
 +  y^2 T_{2,0} + y^3 T_{3,0} + y^4 T_{4,0} + \ldots \right]
 \\
& =  \pi^2 \phi^4 \left[ y \; S_0 (yL) + y^2 \; S_1 (yL) + (K - y^2 \; T_{1,0}) \right],
\end{split}
\label{eq4.2}
\\
\begin{split}
V_{N^2 LL} & =  \pi^2 \phi^4 y \left[ \sum_{n=0}^\infty T_{n,n} (yL)^n + y \sum_{n=1}^\infty T_{n,n-1} (yL)^{n-1} 
 +  y^2 \sum_{n=2}^\infty T_{n,n-2} (yL)^{n-2} + y^3 T_{3,0} + y^4 T_{4,0} + \ldots \right]
 \\
& =  \pi^2 \phi^4 \left[ y S_0 (yL) + y^2 S_1 (yL) + y^3 S_2 (yL) + \left( K - y^2 T_{1,0} - y^3 T_{2,0} \right) \right],
\end{split}
\label{eq4.3}
\\
V_{N^k LL} = \pi^2 \phi^4 \left[ \sum_{p=0}^k y^{p+1} S_p (yL) + K - \sum_{q=2}^{k+1} y^q T_{q-1, 0} \right], \; \; (k \geq 1)
\label{eq4.4}
\end{gather}
Note from Eq.\ (\ref{eq3.4}) for $K$ that the expression (\ref{eq3.17}) for the full SFTP potential is just the $k \rightarrow \infty$ limit of Eq.\ (\ref{eq4.4}):
\begin{equation}
\lim_{k \rightarrow \infty} V_{N^k LL} = \pi^2 \phi^4 \left[\sum_{p=0}^\infty y^{p+1} S_p (yL) + \left(K - \sum_{q=2}^\infty y^q T_{q-1, 0} \right) \right] = V_{SFTP}
\label{eq4.5}
\end{equation}
Indeed, we have chosen to express $K$ by Eq.\  (\ref{eq3.4}) in order to assure the consistency of the limit (\ref{eq4.5}) with $V_{SFTP}$.

We have already seen in the previous section that the series expansion of Eqs.\  (\ref{eq4.1}) and (\ref{eq3.14})
\begin{equation}
V_{LL} = \pi^2 \phi^4 \left[ y + K + 3y^2 L + 9y^3 L^2 + 27 y^4 L^3 + 81y^5 L^4 + \ldots \right]
\label{eq4.6}
\end{equation}
yields the results $y = 0.054135$, $K = -0.058531$, and a running Higgs mass $\left[ V_{LL}^{\prime\prime} (v) \right]^{1/2} = 221.2$ GeV.  The corresponding $NLL$ series expansion of Eq.\  (\ref{eq4.2}), as obtained from series expansions of summations (\ref{eq3.14}) and (\ref{eq3.15}), is given by
\begin{equation}
V_{NLL} = \pi^2 \phi^4 \left[ y + K + B_{NLL} L + C_{NLL} L^2 + D_{NLL} L^3 + E_{NLL} L^4 + \ldots \right]
\label{eq4.7}
\end{equation}
where the $NLL$ numerical value for $y$ is obtained via the series coefficients
\begin{gather}
B_{NLL} (y_{NLL}) = 3y_{NLL}^2 + 3y_{NLL}^3 (4 T_{1,0} - 7) / 2
\label{eq4.8}
\\
C_{NLL} (y_{NLL}) = 9y_{NLL}^3 + (27 T_{1,0} - 621 / 8) y_{NLL}^4
\label{eq4.9}
\\
D_{NLL} (y_{NLL}) = 27 y_{NLL}^4 + 9 (-89 + 24 T_{1,0}) y_{NLL}^5 / 2
\label{eq4.10}
\\
E_{NLL} (y_{NLL}) = 81 y_{NLL}^5 + 27 (240 T_{1,0} - 1049) y_{NLL}^6 / 16 .
\label{eq4.11}
\end{gather}
Substitution of Eq.\  (\ref{eq4.8}) into the condition (\ref{eq2.5}) that ensures minimization of the $NLL$ potential at $\phi = v$ yields the following equation for $T_{1,0}$ in terms of $K$ and the $NLL$ value for $y$:
\begin{equation}
T_{1,0} = - \left( 4K - 21 y_{NLL}^3 + 6y_{NLL}^2 + 4y_{NLL} \right) / 12 y_{NLL}^3 .
\label{eq4.12}
\end{equation}
Similarly, one can substitute Eqs.\  (\ref{eq4.7})--(\ref{eq4.11}) into Eq.\  (\ref{eq2.6}) to find
\begin{equation}
y_{NLL} = \frac{11}{3} B_{NLL} + \frac{35}{3} C_{NLL} + 20 D_{NLL} + 16 E_{NLL}.
\label{eq4.13}
\end{equation}
Given the already-determined numerical value for the aggregate $\phi^4$ counterterm coefficient $K = -0.058531$, we see that Eq.\  (\ref{eq4.12}) can be substituted into each series term (\ref{eq4.8})--(\ref{eq4.11}) within Eq.\  (\ref{eq4.13}) to yield a sixth order polynomial equation for $y_{NLL}$.  One finds only one real positive solution to this equation, $y_{NLL} = 0.053812$ (we discard as spurious a negative real solution $y = -0.176$ as well as complex solutions).  This solution exhibits a controllably small deviation from $y_{LL} = 0.054135$ obtained from the $LL$ approximation to the SFTP potential in the previous section.  We then find from Eq.\  (\ref{eq4.12}) a numerical value for $T_{1,0} = 2.5521$, in which case $B_{NLL}$ (\ref{eq4.8}) and $C_{NLL}$ (\ref{eq4.9}) have numerical values $0.0094373$ and $0.0013294$, respectively. Upon substitution of these $NLL$ numerical values into Eq.\  (\ref{eq2.9}) for the running Higgs boson mass, we find that $\left( V_{NLL}^{\prime\prime} (v) \right)^{1/2} = 227$ GeV, a controllable departure from the $221.2$ GeV $LL$ result of the previous section.

This entire procedure can be repeated for $N^2 LL$, $N^3 LL$ and $N^4 LL$ versions of the SFTP potential.  For the $N^2 LL$ case, one finds from Eq.\  (\ref{eq4.3}) that
\begin{gather}
A_{N^2 LL} (y_{N^2 LL}) = y_{N^2 LL} + K
\label{eq4.14}
\\
B_{N^2 LL} (y_{N^2 LL}) = B_{NLL} (y_{N^2 LL}) 
+y_{N^2 LL}^4\left( 
9T_{2,0}-\frac{81}{4}T_{1,0}+93.9273 
\right)
\label{eq4.15}
\\
C_{N^2 LL} (y_{N^2 LL}) = C_{NLL} (y_{N^2 LL}) + y_{N^2 LL}^5 \left( 54 \; T_{2,0} - \frac{423}{2} \; T_{1,0} + 1001.16 
\right)
\label{eq4.16}
\\
D_{N^2 LL} (y_{N^2 LL}) = D_{NLL} (y_{N^2 LL}) + y_{N^2 LL}^6
\left( 270 T_{2,0}-\frac{5661}{4}T_{1,0} +6872.92 \right) 
\label{eq4.17}
\\
E_{N^2 LL} (y_{N^2 LL}) = E_{NLL} (y_{N^2 LL}) + y_{N^2 LL}^7 
\left(1215 T_{2,0}-\frac{61641}{8} T_{1,0} +38397.6 \right)
\label{eq4.18}
\end{gather}
We find from Eq.\  (\ref{eq2.5}) [{\it i.e.} from $K = -y_{N^2 LL} - B_{N^2 LL} (y_{N^2 LL}) / 2$] that 
\begin{equation}
18 y^4 T_{2,0} = - 4 (y+K) + 6 y^2
+y^3\left( 12\,T_{1,0}-21\right)
 +y^4\left(-\frac{81}{2}T_{1,0}+187.855 \right)
\label{eq4.19}
\end{equation}
where all $y$'s appearing in Eq.\  (\ref{eq4.19}) are understood to be $y_{N^2 LL}$.  Since $\phi^4$-counterterm coefficients $K$ and $T_{1,0}$ have already been numerically determined to be $-0.058531$ and $2.5521$, respectively, and since
\begin{equation}
y_{N^2 LL} = \frac{11}{3} B_{N^2 LL} (y_{N^2 LL}) + \frac{16}{3} C_{N^2 LL} (y_{N^2 LL}) + 20 D_{N^2 LL} (y_{N^2 LL}) + 16 E_{N^2 LL} (y_{N^2 LL})
\label{eq4.20}
\end{equation}
by application of Eq.\ (\ref{eq2.6}) to  the $N^2 LL$ potential (\ref{eq4.3}), we see that 
Eqs.\ (\ref{eq4.19}) and (\ref{eq4.20}) represent two equations in two unknowns:  $y_{N^2 LL}$ and $T_{2,0}$. 
The smallest (and only viable) positive real solution for  $y_{N^2 LL}$ is $0.05392$, in which case $T_{2,0} = -8.1770$.
We then substitute these values, as well as the prior determination of $T_{1,0} = 2.5521$, into $B$ and $C$ within
Eq.\ (\ref{eq2.9}) to find that the $N^2 LL$ running Higgs boson mass is $\left( V_{N^2 LL}^{\prime\prime} (v) \right)^{1/2} = 225.0$ GeV. Thus it is clear that $y_{N^2 LL}$ and the $N^2 LL$ Higgs boson mass are controllable departures from their $LL$ counterparts.

For completeness, we list below the corresponding results for $N^3 LL$ and $N^4 LL$ version of the SFTP potential as obtained from series solutions to Eqs.\  (\ref{eq3.12}) and (\ref{eq3.13}):
{\allowdisplaybreaks
\begin{gather}
B_{N^3 LL} = B_{N^2 LL} + y^5 
\left(12\,T_{3,0}-30\, T_{2,0}+186.730\, T_{1,0}-1352.65  \right)
\label{eq4.21}
\\
C_{N^3 LL} = C_{N^2 LL} + y^6 \left( 90 \; T_{3,0} - \frac{3231}{8} \; T_{2,0} 
+2641.53 \, T_{1,0}-17523.9\right)
\label{eq4.22}
\\
D_{N^3 LL}  =  D_{N^2 LL}
 + y^7 
 \left(
 540 \, T_{3,0} -\frac{13257}{4}\, T_{2,0} +22690.0 \, T_{1,0}-143.417\times 10^3 
  \right) 
 \label{eq4.23}
 \\  
E_{N^3 LL}  =  E_{N^2 LL}
 + y^8 
 \left(
 2835 \, T_{3,0} -\frac{342387}{16}\, T_{2,0}+ 152.644\times 10^3-937.659\times 10^3
  \right) 
\label{eq4.24}
\\
B_{N^4 LL} -B_{N^3 LL} =    y^6 \left( 
15 \; T_{4,0} -\frac{159}{4}\; T_{3,0}+279.532\; T_{2,0} 
-2696.44 \; T_{1,0} +24032.2
\right) ,
\label{eq4.25}
\\
C_{N^4 LL}-C_{N^3 LL}  =    y^7 \left( 135 \; T_{4,0} - \frac{2619}{4} \; T_{3,0}  
+ 4933.79 \; T_{2,0}  +44780.1\; T_{1,0}+360.414\times 10^3
\right),
\label{eq4.26}
\\
D_{N^4 LL}-D_{N^3 LL}  =   y^8 \left( 
945\, T_{4,0} -\frac{25443}{4}\, T_{3,0} +50885.5\, T_{2,0}
-446.879\times 10^3\,T_{1,0}+337.923\times 10^4
  \right)  ,
\label{eq4.27}
\\
\begin{split}
E_{N^4 LL}-E_{N^3 LL}   =&  y^9\Biggl(
5670\, T_{4,0} -\frac{94959}{2}\, T_{3,0}+400.145\times 10^3\, T_{2,0}\Biggr.
\\
&\qquad\qquad\qquad-\Biggl.345.173\times 10^4T_{1,0}+250.285\times 10^5
\Biggr)  .
\end{split}
\label{eq4.28}
\end{gather}
}

At the $N^3 LL$ level, the condition $K = -y_{N^3 LL} - B_{N^3 LL} (y_{N^3 LL}) / 2$ implies that
\begin{equation}
\begin{split}
24 y^5\, T_{3,0}  =& 
- 4 (y+K) + 6 y^2+y^3\left( 12\,T_{1,0}-21\right)
 +y^4\left(-\frac{81}{2}T_{1,0}+187.855 \right)
 \\
&+y^5\left(
60 \; T_{2,0}-373.459\; T_{1,0}+2705.30
 \right)~,
\end{split}
\label{eq4.29}
\end{equation}
where $y = y_{N^3 LL}$.  Given the previously determined numerical values $K = -0.058531, \; T_{1,0} = 2.5521, \; T_{2,0} = -8.1770$, and the constraint (\ref{eq2.6}) for $y_{N^3 LL}$ in terms of $\{B, C, D, E \}_{N^3 LL}$, we find that the smallest positive real root for $y_{N^3 LL}$ is $y_{N^3 LL} = 0.05385$, in which case $T_{3,0} = 83.211$ and $\left[ V_{N^3 LL}^{\prime\prime} (v) \right]^{1/2} = 226$ GeV.

At the $N^4 LL$ level, the condition $K = -y_{N^4 LL} - B_{N^4 LL} (y_{N^4 LL}) / 2$ implies that
\begin{equation}
\begin{split}
30 y^6\, T_{4,0}  =& 
- 4 (y+K) + 6 y^2+y^3\left( 12\,T_{1,0}-21\right)
 +y^4\left(-\frac{81}{2}T_{1,0}+187.855 \right)
 \\
&+y^5\left(
60 \; T_{2,0}-373.459\; T_{1,0}+2705.30
 \right)
 \\
 &+y^6\left(
 \frac{159}{2}\, T_{3,0}-559.064\, T_{2,0}+5392.87\, T_{1,0}-48064.4
 \right) ~,
\end{split}
\label{eq4.30}
\end{equation}
where $y=y_{N^4 LL}$.  Upon substituting $K = -0.058531, \; T_{1,0} = 2.5521, \; T_{2,0} = -8.1770, \; T_{3,0} = 83.211$ into Eqs.\ (\ref{eq4.21})--(\ref{eq4.28}), we then find from Eq.\ (\ref{eq2.6}) that the only sufficiently small positive root (in fact, the smallest positive root) for $y_{N^4 LL}$ is $y_{N^4 LL} = 0.05391$.  We then find from Eq.\ (\ref{eq4.26}) that $T_{4,0} = -1141.8$, and that $\left[ V_{N^4 LL}^{\prime\prime}(v)\right]^{1/2} = 225$ GeV.

These results for the SFTP potential are summarized in Table \ref{vtab1}.  There is striking order-by-order stability in the predictions obtained for $y \; (\cong 0.054)$ and for the running Higgs boson (221--227 GeV).  In the next section we will demonstrate how these predictions are only minimally altered by ``turning on'' the $t$-quark Yukawa couplant $x$ and the QCD couplant $z$ to their physical values (\ref{eq2.1}).

\begin{table}[hbt]
\centering
\begin{tabular}{||c|c|c|c||}     \hline\hline
 $k$ & $y$ & $T_{k,0}$ & $\left[ V^{\prime\prime}(v)\right]^{1/2}$ \\ \hline\hline
0  & 0.05414 & 1 & 221.2 \\ \hline
1 & 0.05381 & 2.552 & 227.0   \\ \hline
2  & 0.05392 & -8.117 & 224.8 \\ \hline
3 & 0.05385 & 83.21 & 226.3  \\ \hline
4 & 0.05391 & -1142  & 225.0\\ \hline\hline
\end{tabular}
\caption{Results for the SFTP potential taken to $N^4 LL$ order, as discussed in Section 4.  $T_{k,0} \; (k \geq 1)$ is the coefficient of the finite $\pi^2 \phi^4 y^{k+1}$ counterterm contributing to the SFTP potential. The final column is in GeV units.}
\label{vtab1}
\end{table}

To conclude this Section, we consider the impact of a perturbation from a conventional-symmetry breaking quadratic term on the above predictions.  This additional tree-level contribution to the effective potential has the form
\begin{equation}
\Delta V^{(2)}=-\frac{1}{2}M^2\phi^2~,
\end{equation}
and as noted in \cite{1,3}, this term generates an imaginary part in the effective potential for small values of $\phi$ at one-loop order.  However, near the electroweak scale $\phi=v$ [the scale at which the effective potential has a minimum (\ref{eq2.5}), renormalization conditions are applied (\ref{eq2.6}), and the Higgs mass is defined (\ref{eq2.9})], this quadratic term can be treated as a perturbation.  Expanding the one-loop results of \cite{3} to leading order in $\epsilon=M/v$  results in the following perturbation to the effective potential
\begin{equation}
\Delta V_{eff}^{(2)}=-\frac{1}{2}\epsilon^2 v^2\phi^2+\frac{1}{16}y \epsilon^2 v^2\phi^2
\left[-6\left(1+\log 3\right)-12\log{\left(4\pi^2 y\phi^2/v^2\right)}  \right]~.
\label{epsilon_exp}
\end{equation}
We note that implicit in this expansion is the restriction $M^2\ll \lambda v^2$ arising from expansion of logarithmic terms such as $\log{\left(-M^2+\lambda\phi^2\right)}$, which clearly excludes the conventional-symmetry-breaking (CSB) regime $M^2= \lambda v^2$, and thus (\ref{epsilon_exp}) truly represents a perturbation to the radiative scenario from a CSB mass term. Combining (\ref{epsilon_exp}) with the SFTP leading-logarithm results allows us to examine the numerical effect of the perturbation on the leading-logarithm ($k=0$) results of Table \ref{vtab1}.  For $\epsilon=0.1$ ({\it i.e.,} $M\approx 25\,{\rm GeV}$), $y$  decreases slightly to $y=0.05392$ and $m_H$ increases marginally to $m_H=223.4\,{\rm GeV}$.  Even for $\epsilon=0.2$ ($M\approx 50\,{\rm GeV}$), which results in $y=0.05328$ and $m_H=229.8\,{\rm GeV}$, the  
effect $\Delta m_H\sim 10\,{\rm GeV}$ on the Higgs mass is still far less than the CSB expectation $\Delta m_H\sim M\approx 50\,{\rm GeV}$.  We thus conclude that the $LL$ radiatively-broken SFTP scenario is not destabilized by a perturbation from a CSB mass term in the effective potential.   



\section{Turning on the Yukawa Sector}

\renewcommand{\theequation}{5.\arabic{equation}}
\setcounter{equation}{0}

The subdominant electroweak couplants $x$ and $z$ (\ref{eq2.1}) provide the largest alteration to the SFTP potential.  If $x = 0$ ({\it i.e.,} if there is no Yukawa coupling), then there is no way for $z$, the QCD quark-gluon coupling, to enter the potential.  Diagrammatically, $z$ enters the potential beginning at two-loop order as a virtual gluon exchange within a $t$-quark loop.

\begin{figure}[hbt]
\centering
\includegraphics[scale=0.8]{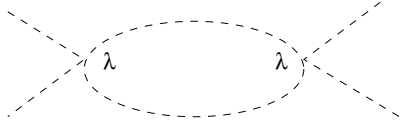}
\caption{The divergent 1-loop graph leading to the finite $\pi^2 T_{1,0} y^2 \phi^4$ counterterm $\left( T_{1,0} = D_{0,2,0,0} \right)$}
\label{fig1}
\end{figure}

\begin{figure}[hbt]
\centering
\includegraphics[scale=0.8]{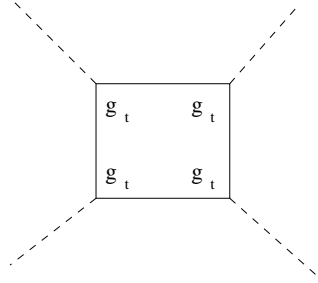}
 \caption{The divergent 1-loop graph leading to the finite $\pi^2 U x^2 \phi^4$ counter-term 
 $\left( U = D_{2,0,0,0} \right)$.}
\label{fig2}
\end{figure}

If we augment the SFTP potential with contributions from the subdominant electroweak couplants $x$ and $z$, the potential one obtains [in the absence of ($u, d, s, c, b$) Yukawa couplings and electroweak gauge couplants] is just
\begin{equation}
V_{xyz} = \pi^2 \phi^4 \sum_{n=0}^\infty x^n \sum_{k=0}^\infty y^k \sum_{\ell=0}^\infty z^\ell \sum_{p=0}^{n+k+\ell-1} L^p \; D_{n,k,\ell,p} = \pi^2 \phi^4 S
\label{eq5.1}
\end{equation}
where $D_{0,0,0,0} = D_{1,0,0,0} = D_{0,0,1,0} = 0, \; \; D_{0,1,0,0} = 1$.  The leading logarithm $(LL)$ portion of $S$ is comprised of those terms in Eq.\  (\ref{eq5.1}) for which $p = n + k + \ell - 1$.  Thus coefficients $C_{n,k,\ell}$ in Eq.\  (\ref{eq2.2}) are just coefficients $D_{n,k,\ell,n+k+\ell - 1}$ in Eq.\ (\ref{eq5.1}).  Similarly the next-to-leading logarithm $(NLL)$ portion of $S$ is comprised of those terms in Eq.\  (\ref{eq5.1}) for which $p = n + k + \ell - 2$.  The series $S$ may be expressed as a power series in $L \equiv \log (\phi^2/v^2)$ of the form (\ref{eq2.4}):
\begin{equation}
S = (y + K) + BL + CL^2 + DL^3 + EL^4 + \ldots
\label{eq5.2}
\end{equation}
where $\{B_{LL}, C_{LL}, D_{LL}, E_{LL} \}$, as obtained from the RGE (\ref{eq2.3}) with one-loop $RG$ functions, are given by Eqs.\  (7.2)--(7.5) of ref.\ \cite{6}.  As in the SFTP case, the constant $K$ in (\ref{eq5.2}) corresponds to the aggregate contribution of purely-$\phi^4$ finite terms contributing to the potential (\ref{eq5.1}) after infinities are removed:
\begin{equation}
K = \sum_{p=2}^\infty \sum_{n=0}^p \sum_{k=0}^{p-n} x^n y^k z^{p-n-k} D_{n, k, p-n-k, 0} .
\label{eq5.3}
\end{equation}
Thus the SFTP counterterm coefficients $T_{k,0}$ in the previous section correspond to coefficients $D_{0,k,0,0}$ in Eq.\  (\ref{eq5.3}).  However, Eq.\  (\ref{eq5.3}) now includes terms not present in the SFTP version of $K$ given by Eq.\  (\ref{eq3.4});  Eq.\  (\ref{eq3.4}) is just the sum of the $x = z = 0$ subset of terms contributing to Eq.\  (\ref{eq5.3}).  Recall from Section 2 that $K$ was found numerically via Eqs.\  (\ref{eq2.5}) and (\ref{eq2.6}) to be equal to $-0.057935$ \cite{5} from the leading logarithm version of the potential (\ref{eq5.1}).  This is only a small departure from the value $K = -0.058531$ obtained from the SFTP potential (\ref{eq3.1}).  This difference reflects the presence of additional finite $\phi^4$ terms.  For example, the $\pi^2 \phi^4 y^2 T_{1,0} \; (= \pi^2 \phi^4 y^2 D_{0,2,0,0})$ term associated with the post-subtraction contribution of the 1-loop divergent graph of Fig.\ \ref{fig1} is now augmented by a $\pi^2 \phi^4 x^2 D_{2,0,0,0} \; (\equiv \pi^2 \phi^4 x^2 U)$ term associated with the post-subtraction contribution of the 1-loop divergent graph of Fig.\ \ref{fig2}.

As in the SFTP case, we approximate the full potential (\ref{eq5.1}) via a series of successive approximations:
{\allowdisplaybreaks 
\begin{gather}
V_{LL} = \pi^2 \phi^4 \left[ \left( \sum_{p=1}^\infty \sum_{n=0}^p \sum_{k=0}^{p-n} x^n y^k z^{p-n-k} L^{p-1} D_{n,k,p-n-k, p-1} \right) + K \right],
\label{eq5.4}
\\
\begin{split}
V_{NLL}  = \pi^2 \phi^4 &\left[ \left( \sum_{p=1}^\infty \sum_{n=0}^p \sum_{k=0}^{p-n} x^n y^k z^{p-n-k} L^{p-1} D_{n,k,p-n-k, p-1} \right)\right.
\\
& +  \left(\sum_{p=2}^\infty \sum_{n=0}^p \sum_{k=0}^{p-n} x^n y^k z^{p-n-k} L^{p-2} D_{n,k,p-n-k, p-2} \right)
\\
& +  \left. K - \left(\sum_{n=0}^2 \sum_{k=0}^{2-n} x^n y^k z^{2-n-k} D_{n,k,2-n-k,0} \right)\right]
\end{split}
\label{eq5.5}
\\
\begin{split}
V_{N^q LL}  = & \pi^2 \phi^4 \left[ \left( \sum_{m=0}^q \sum_{p = m+1}^\infty \sum_{n=0}^p \sum_{k=0}^{p-n} x^n y^k z^{p-n-k} L^{p-m-1} D_{n,k,p-n-k, p-m-1} \right)\right.
\\
& +  K - \left. \left( \sum_{p=2}^{q+1} \sum_{n=0}^p \sum_{k=0}^{p-n} x^n y^k z^{p-n-k} D_{n,k, p-n-k, 0} \right) \right], \; \; (q \geq 1),
\end{split}
\label{eq5.6}
\end{gather}
}
where the final expression in Eq.\  (\ref{eq5.5}) is just the two finite 1-loop $\phi^4$ post-subtraction terms associated with Figs.~\ref{fig1} and \ref{fig2}:
\begin{equation}
\sum_{n=0}^2 \sum_{k=0}^{2-n} x^n y^k z^{p-n-k} D_{n, k, 2-n-k, 0} = T_{1,0} y^2 + U x^2.
\label{eq5.7}
\end{equation}
Comparison of Eq.\  (\ref{eq5.6}) to Eq.\  (\ref{eq5.1}) through use of the definition (\ref{eq5.3}) for $K$ shows that the full potential $V_{xyz}$ is just the $q \rightarrow \infty$ limit of $V_{N^q LL} : \lim_{q\rightarrow \infty} V_{N^q LL} = V_{xyz}$, analogous to Eq.\  (\ref{eq4.5}) for SFTP case.

We now use the two loop RG functions to obtain the $NLL$ contributions to the series $S$ in Eq.\  (\ref{eq5.1}). To do this, we break these $\overline{MS}$ functions up into their known one-loop (1L) and two-loop (2L) components \cite{11}:
\begin{gather}
\gamma = \gamma_{1L} + \gamma_{2L}, \; \; \gamma_{1L} = 3x/4, \; \; \gamma_{2L} = \frac{3y^2}{8} - \frac{27x^2}{64} + \frac{5xz}{4};
\label{eq5.8}
\\
\beta_x = \beta_{x1L} + \beta_{x2L}, \; \; \beta_{x1L} = \frac{9}{4} x^2 - 4xz,
\nonumber\\  
\beta_{x2L} = - \frac{3}{2} x^3 - \frac{3}{2} x^2 y + \frac{3}{4} xy^2 + \frac{9}{2} x^2 z - \frac{27}{2} x z^2 ;
\label{eq5.9}
\\
\beta_y = \beta_{y1L} + \beta_{y2L}, \; \; \beta_{y1L} = 6y^2 + 3yx - \frac{3}{2} x^2,
\nonumber\\
\beta_{y2L} = -\frac{39}{2} y^3 - 9xy^2 - \frac{3}{16} x^2 y + \frac{15}{8} x^3 + 5xyz - 2x^2 z ;
\label{eq5.10}
\\
\beta_z = \beta_{z1L} + \beta_{z2L}, \; \; \beta_{z1L} = -\frac{7}{2} z^2, \; \; \beta_{z2L} = -\frac{13}{4} z^3 - \frac{xz^2}{4}.
\label{eq5.11}
\end{gather}
We also break up the series coefficients in Eq.\  (\ref{eq5.2}) into their $LL$ and $NLL$ contributions, 
\begin{gather}
A \equiv y + K = y + \left( T_{1,0} y^2 + U x^2\right) + \ldots
\label{eq5.12}
\\
B = B_{LL} + \Delta B_{NLL} + \ldots
\label{eq5.13}
\\
C = C_{LL} + \Delta C_{NLL} + \ldots
\label{eq5.14}
\\
D = D_{LL} + \Delta D_{NLL} + \ldots
\label{eq5.15}
\\
E = E_{LL} + \Delta E_{NLL} + \ldots
\label{eq5.16}
\end{gather}
If we substitute the series (\ref{eq5.2}) into the RGE (\ref{eq2.3}) using the RG functions (\ref{eq5.8})--(\ref{eq5.11}), we find that the cancellation of ${\cal{O}}(L^0)$ terms on the left hand side of (\ref{eq2.3}) implies that
\begin{gather}
2 B_{LL} = \beta_{y1L} - 4 \gamma_{1L}\; y.
\label{eq5.17}
\\
2 \Delta B_{NLL} = -2\gamma_{1L} B_{LL} + 2 \beta_{x1L} Ux + 2 \beta_{y1L}  T_{1,0} y
+ \beta_{y2L} - 4\gamma_{1L} \; (T_{1,0} y^2 + U x^2) - 4\gamma_{2L}\; y.
\label{eq5.18}
\end{gather}
We see that from Eq.\  (\ref{eq5.17}) that
\begin{equation}
B_{LL} = 3y^2 - 3x^2/4,
\label{eq5.19}
\end{equation}
consistent with the effective potential of Coleman and Weinberg \cite{1}.  Upon substituting Eq.\  (\ref{eq5.19}) into (\ref{eq5.18}) we find that
\begin{equation}
\Delta B_{NLL} = \left( -\frac{27}{4} + \frac{3 T_{1,0}}{2} \right) xy^2 + \left( \frac{3}{2} + \frac{3U}{4} \right) x^3 - (1+4U) x^2 z 
+ \left( - \frac{21}{2} + 6 T_{1,0} \right) y^3 + \left( \frac{3}{4} - \frac{3T_{1,0}}{2} \right) x^2 y.
\label{eq5.20}
\end{equation}

For ${\cal{O}}(L^1)$ terms of the RGE (\ref{eq2.3}) to vanish, we must have
\begin{equation}
4CL = - 4\gamma CL + \beta_x \frac{\partial B}{\partial x} L + \beta_y \frac{\partial B}{\partial y} L + \beta_z \frac{\partial B}{\partial z} L - 4\gamma B L,
\label{eq5.21}
\end{equation}
hence that
\begin{equation}
4C_{LL} = \beta_{x1L} \frac{\partial B_{LL}}{\partial x} + \beta_{y1L} \frac{\partial B_{LL}}{\partial y} + \beta_{z1L} \frac{\partial B_{LL}}{\partial z} - 4 \gamma_{1L} B_{LL}
\label{eq5.22}
\end{equation}
and that
\begin{equation}
\begin{split}
4 \Delta C_{NLL}  = & -4 \gamma_{1L} C_{LL} + \beta_{x1L} \frac{\partial \Delta B_{NLL}}{\partial x} + \beta_{x2L} \frac{\partial B_{LL}}{\partial x}
 +  \beta_{y1L} \frac{\partial \Delta B_{NLL}}{\partial y} 
 + \beta_{y2L} \frac{\partial B_{LL}}{\partial y}
 \\
& +  \beta_{z1L} \frac{\partial \Delta B_{NLL}}{\partial z} + \beta_{z2L} \frac{\partial B_{LL}}{\partial z}
 -  4 \gamma_{1L} \Delta B_{NLL} - 4 \gamma_{2L} B_{LL}.
\end{split}
\label{eq5.23}
\end{equation}
Given our solution (\ref{eq5.19}) for $B_{LL}$ and the $1L$ RG functions, one recovers the result \cite{5,6}
\begin{equation}
C_{LL} = 9y^3 + \frac{9}{4} xy^2 - \frac{9}{4} x^2 y + \frac{3}{2} x^2 z - \frac{9}{32} x^3.
\label{eq5.24}
\end{equation}
Given our solution (\ref{eq5.20}) for $\Delta B_{NLL}$ and the $1L$ and $2L$ RG functions in Eqs.\  (\ref{eq5.8})--(\ref{eq5.11}), we find that
\begin{equation}
\begin{split}
\Delta C_{NLL}  = & \left( -\frac{621}{8} + 27 T_{1,0} \right) y^4 + \left( \frac{27 T_{1,0}}{2} - \frac{225}{4} \right) xy^3
 +  \left(-\frac{3 T_{1,0}}{2} 
+ \frac{21}{2} \right) xy^2 z 
\\&+ \left( 3 T_{1,0} - \frac{9}{2} \right) x^2 yz
 +  \left( -\frac{225 T_{1,0}}{32} + \frac{27}{8} \right) x^2 y^2 
+ \left( \frac{23 U}{2} + \frac{127}{16} \right) x^2 z^2
\\
& +  \left( -\frac{27}{4} - \frac{15 U}{4}\right) x^3 z 
 + \left( -\frac{45 T_{1,0}}{16} + \frac{351}{32} \right) x^3 y
 +  \left( \frac{405}{256} + \frac{45 U}{64} + \frac{9 T_{1,0}}{16}\right) x^4.
\end{split}
\label{eq5.25}
\end{equation}
The ${\cal{O}}(L^2)$ terms in the RGE (\ref{eq2.3}) vanish provided
\begin{equation}
6 DL^2  =  -6 \gamma DL^2 + \beta_x \frac{\partial C}{\partial x} L^2 + \beta_y \frac{\partial C}{\partial y} L^2
 + \beta_z \frac{\partial C}{\partial z} L^2 - 4 \gamma CL^2.
\label{eq5.26}
\end{equation}
Terms degree-4 in couplants cancel in Eq.\  (\ref{eq5.26}) provided
\begin{equation}
6 D_{LL} = \beta_{x1L} \frac{\partial C_{LL}} {\partial x} + \beta_{y1L} \frac{\partial C_{LL}}{\partial y} + \beta_{z1L} \frac{\partial C_{LL}}{\partial z} - 4 \gamma_{1L} C_{LL}.
\label{eq5.27}
\end{equation}
and terms degree-5 in couplants cancel provided
\begin{equation}
\begin{split}
6 \Delta D_{NLL}  = & - 6\gamma_{1L} D_{LL} + \beta_{x1L} \frac{\partial \Delta C_{NLL}}{\partial x} + \beta_{x2L} \frac{\partial C_{LL}}{\partial x} 
 +  \beta_{y1L} \frac{\partial \Delta C_{NLL}}{\partial y} 
 \\
 &+ \beta_{y2L} \frac{\partial C_{LL}}{\partial y} 
 + \beta_{z1L} \frac{\partial \Delta C_{NLL}}{\partial z}
 +  \beta_{z2L}\frac{\partial C_{LL}}{\partial z} - 4 \gamma_{1L} \Delta C_{NLL} - 4 \gamma_{2L} C_{LL}.
\end{split}
\label{eq5.28}
\end{equation}
If one substitutes $C_{LL}$ [Eq.\  (\ref{eq5.24})] into (\ref{eq5.27}), one recovers Eq.\  (7.4) of ref.\ \cite{6}:
\begin{equation}
D_{LL}  =  27y^4 + \frac{27}{2} xy^3 - \frac{3}{2} xy^2 z + 3x^2yz  
 -  \frac{225}{32} x^2 y^2 - \frac{23}{8} x^2 z^2 + \frac{15}{16} x^3 z 
 -  \frac{45}{16} x^3 y + \frac{99}{256} x^4.
 \label{eq5.29}
\end{equation}
Similarly, substitution of $C_{LL}$ and $\Delta C_{NLL}$ [Eqs.\  (\ref{eq5.24}) and (\ref{eq5.25})] into Eq.\  (\ref{eq5.28}) determines $\Delta D_{NLL}$:
\begin{equation}
\begin{split}
\Delta D_{NLL}  = & \left( -\frac{801}{2} + 108 T_{1,0} \right) y^5 + \left( -\frac{11547}{32} + 81 T_{1,0} \right) xy^4
+ \left( -12 T_{1,0} + \frac{147}{2} \right) xy^3 z 
\\
& +  \left( \frac{15 T_{1,0}}{8} - \frac{291}{16} \right) xy^2 z^2 + \left( - \frac{23 T_{1,0}}{4}+ \frac{75}{4} \right) x^2 yz^2 + \left( -\frac{45}{32} - \frac{45 T_{1,0}}{2} \right) x^2 y^3
\\
&  +  \left( \frac{177 T_{1,0}}{16} - \frac{33}{8} \right) x^2 y^2 z 
+ \left( - \frac{877}{32} - \frac{115 U}{4} \right) x^2 z^3 + \left( \frac{3125}{128} + \frac{201 U}{16} \right) x^3 z^2 
\\
& +  \left( \frac{69 T_{1,0}}{8} - \frac{615}{16} \right) x^3 yz + \left( \frac{19323}{256} - \frac{2781 T_{1,0}}{128} \right) x^3 y^2 + \left( -\frac{1023}{128} - \frac{135 U}{32} - \frac{9 T_{1,0}}{4} \right) x^4 z
\\
& +  \left( \frac{3825}{256} + \frac{45 T_{1,0}}{128} \right) x^4 y + \left( -\frac{1035}{512} + \frac{45 U}{64} + \frac{81 T_{1,0}}{64} \right) x^5
\end{split}
\label{eq5.30}
\end{equation}
Finally, to cancel ${\cal{O}}(L^3)$ terms in the RGE (\ref{eq2.3}), we see that terms degree-5 in couplants cancel provided
\begin{equation}
8E_{LL} = \beta_{x1L} \frac{\partial D_{LL}}{\partial x} + \beta_{y1L} \frac{\partial D_{LL}}{\partial y} + \beta_{z1L} \frac{\partial D_{LL}}{\partial z} - 4\gamma_{1L} D_{LL},
\label{eq5.31}
\end{equation}
and that terms degree-6 in couplants cancel provided
\begin{equation}
\begin{split}
8 \Delta E_{NLL}  = & -8 \gamma_{1L} E_{LL} + \beta_{x1L} \frac{\partial \Delta D_{NLL}}{\partial x}
 +  \beta_{x2L} \frac{\partial D_{LL}}{\partial x} 
 + \beta_{y1L} \frac{\partial \Delta D_{NLL}}{\partial y} 
\\
&+ \beta_{y2L} \frac{\partial D_{LL}}{\partial y}
 +  \beta_{z1L} \frac{\partial \Delta D_{NLL}}{\partial z} 
 + \beta_{z2L} \frac{\partial D_{LL}}{\partial z}
 -  4 \gamma_{1L} \Delta D_{NLL} - 4 \gamma_{2L} D_{LL}.
\end{split}
\label{eq5.32}
\end{equation}
Upon substitution of Eq.\  (\ref{eq5.29}) into Eq.\  (\ref{eq5.31}), we recover Eq.\  (7.5) of 
ref.\ \cite{6}:
\begin{equation}
\begin{split}
E_{LL}  = & 81y^5 + \frac{243}{4} xy^4 - 9xy^3 z + \frac{45}{32} xy^2 z^2
 -  \frac{69}{16} x^2 yz^2 - \frac{135}{8} x^2 y^3 + \frac{531}{64} x^2 y^2 z
 \\
& +  \frac{345}{64} x^2 z^3 - \frac{603}{256} x^3 z^2 + \frac{207}{32} x^3 yz
 -  \frac{8343}{512} x^3 y^2 - \frac{459}{512} x^4 z + \frac{135}{512} x^4 y + \frac{837}{1024} x^5.
\end{split}
\label{eq5.33}
\end{equation}
Substitution of Eqs.\  (\ref{eq5.29}) and (\ref{eq5.30}) into Eq.\  (\ref{eq5.32}) yields the corresponding $NLL$ contribution to the series (\ref{eq5.2}):
\begin{equation}
\begin{split}
\Delta E_{NLL}  = & \left( - \frac{55539 T_{1,0}}{4096} + \frac{1081377}{8192} \right) y^2 x^4 + \left( \frac{2187 T_{1,0}}{256} - \frac{29133}{2048} \right)yx^5 
 + \left( - \frac{125793}{64} + 405 T_{1,0} \right) y^5 x 
 \\
& +  \left( \frac{1035 U}{64} + \frac{105 T_{1,0}}{16} + \frac{111633}{4096} \right) x^4 z^2 + \left( - \frac{1215 T_{1,0}}{32} - \frac{195939}{1024} \right) y^4 x^2
\\
& +  \left( \frac{4255 U}{64} + \frac{38613}{512} \right) x^2 z^4 + \left( -\frac{315 U}{64} - \frac{207 T_{1,0}}{32} + \frac{17703}{2048} \right) x^5 z 
\\
& +  \left( -\frac{4509 U}{128} - \frac{75315}{1024} \right) x^3 z^3 + \left( - \frac{28323}{16} + 405 T_{1,0} \right) y^6 + \left( \frac{4581}{64} + \frac{855 T_{1,0}}{32}\right) y^3 x^2 z
\\
& +  \left( - \frac{31455 T_{1,0}}{256} + \frac{227529}{512} \right) y^3 x^3 + \left( \frac{3231}{8} - \frac{135 T_{1,0}}{2} \right) y^4 xz
\\
& +  \left( -\frac{3807}{32} + \frac{225 T_{1,0}}{16}\right) y^3 xz^2 + \left( -\frac{28197}{128} + \frac{7191 T_{1,0}}{128} \right) y^2 x^3 z + \left( \frac{14847}{512} - \frac{1215 T_{1,0}}{64} \right) y^2 x^2 z^2 
\\
& +  \left( \frac{3603}{128} - \frac{165 T_{1,0}}{64}\right) y^2 xz^3 + \left( - \frac{35145}{512} + \frac{621 T_{1,0}}{256}\right) yx^4 z + \left( \frac{7323}{64}- \frac{2643 T_{1,0}}{128}\right) yx^3 z^2 \\
& +  \left( -\frac{93}{2} + \frac{345 T_{1,0}}{32} \right) yx^2 z^3 
+ \left( - \frac{208629}{32768} + \frac{1485 U}{2048} + \frac{1269 T_{1,0}}{1024} \right) x^6
\end{split}
\label{eq5.34}
\end{equation}
One could continue this procedure indefinitely to obtain  ${\cal{O}}(L^5)$, ${\cal{O}}(L^6)$, {\it etc} 
NLL contributions to the series (\ref{eq5.2}); thus, one can in principle obtain the entire NLL contribution to the full potential $V_{xyz}$ (\ref{eq5.1}). However, we have already seen
that the conditions (\ref{eq2.5}) and (\ref{eq2.6}) are sensitive only up to ${\cal{O}}(L^4)$ terms in the potential series $S$.



\section{Yukawa Sector Corrections to SFTP Results}

\renewcommand{\theequation}{6.\arabic{equation}}
\setcounter{equation}{0}

Since the scalar field couplant $y (=\lambda / 4\pi^2)$ is the dominant couplant of the $SM$, with the $t$-quark Yukawa couplant $x (=g_t^2 / 4\pi^2)$ and the QCD couplant $z (=g_3^2 / 4\pi^2 = \alpha_s / \pi)$ characterizing the less dominant Yukawa sector of the effective potential [$z$ does not enter the effective potential unless $x$ is nonzero], one test of order-by-order stability of the effective potential is to augment the SFTP of Section 4 with $LL$ contributions from the Yukawa sector.  These are given explicitly by Eqs.\ (\ref{eq5.19}), (\ref{eq5.24}), (\ref{eq5.27}) and (\ref{eq5.33}) of the previous section.  Thus, we assume here that
\begin{equation}
	\left( 
		\begin{array}{c}
			B\\C\\D\\E
		\end{array}
	\right)_{y_{N^k LL} + \{x,z\}_{LL}}
=
	\left(
		\begin{array}{c}
			B\\C\\D\\E
		\end{array}
	\right)_{\{x,y,z\}_{LL}}
+
	\left(
		\begin{array}{c}
			\Delta B \\ \Delta C \\ \Delta D \\ \Delta E
		\end{array}
	\right)_{N^k LL \; SFTP}
\label{eq6.1}
\end{equation}

For example, the augmentation of the $NLL$ SFTP with $LL$ contributions from the Yukawa Sector leads to the following series coefficients in Eq.\ (\ref{eq2.6})
{\allowdisplaybreaks
\begin{gather}
B(y,T_{1,0}) = 3y^2 - \frac{3x^2}{4} + \left(6 T_{1,0} - 21/2\right) y^3,
\label{eq6.2}
\\
C(y, T_{1,0})  =  9y^3 + \frac{9}{4} xy^2 - \frac{9}{4} x^2 y + \frac{3}{2} x^2 z - \frac{9}{32} x^3 
 +  (27 T_{1,0} - 621/8) y^4,
\label{eq6.3}
\\
\begin{split}
D(y,T_{1,0})  =&  27y^4 + 27 xy^3/2 - 3xy^2 z / 2 - 225 x^2 y^2 / 32
 -  23x^2 z^2 / 8 + 15x^3 z / 16 
\\
&- 45 x^3 y / 16
 +  99 x^4 / 256 + (108 T_{1,0} - 801 / 2) y^5
\end{split}
\label{eq6.4}
\\
\begin{split}
E(y,T_{1,0})  = & 81y^5 + 243 xy^4/4 - 9xy^3 z + 45 xy^2 z^2 / 32
 -  69 x^2 yz^2 / 16 - 135 x^2 y^3 / 8 + 531 x^2 y^2 z / 64 
\\
& +  345 x^2 z^3 / 64 - 603 x^3 z^2 / 256
 +  207 x^3 yz / 32 - 8343x^3 y^2 / 512
\\
& -  459 x^4 z / 512 + 135 x^4 y / 512 + 837 x^5 / 1024
 +  (405 T_{1,0} - 28323/16) y^6,
\end{split}
\label{eq6.5}
\end{gather}
}
The conditions (\ref{eq2.5}) and (\ref{eq2.6}) in conjunction with the prior determination of $K = -0.057935$ (see Sec. 2) enable one to have two equations in the two unknowns $T_{1,0}$ and $y$ (or equivalently $T_{1,0} y^3$ and $y$),
\begin{gather}
T_{1,0} y^3 = - \frac{(y+K)}{3} - \frac{y^2}{2} + \frac{x^2}{8} + \frac{7y^3}{4},
\label{eq6.6}
\\
y  =  \frac{11}{3} B(y, T_{1,0}) + \frac{35}{3} C(y, T_{1,0})
 +  20 D(y, T_{1,0}) + 16 E (y, T_{1,0}).
\label{eq6.7}
\end{gather} 
Utilizing the values $x(v) = 0.0253$, $z(v) = 0.0329$, $K = -0.05793$, one finds that $y(v) = 0.05351$ and hence that $T_{1,0} = 2.5533$.  Note that these values are only small departures from the $NLL \;$ $SFTP$, consistent with $y$ being the dominant $SM$ couplant.  One then finds from Eq.\ (\ref{eq2.9}) that $\left[ V^{\prime\prime} (v)\right]^{1/2} = 222$ GeV, a $5$ GeV decrease from the $NLL \; \; SFTP$ value of Table \ref{vtab1}.

One can, of course, continue this procedure to $k = \{ 2,3,4 \}$ levels in Eq.\ (\ref{eq6.1}).  For example, if $k = 2$, the finite counterterm coefficient $T_{2,0}$ is obtainable from the minimization condition $K = -B/2 - y$, where from Eq.\ (\ref{eq4.15}),
\begin{equation}
B(y, T_{1,0}, T_{2,0})  =  3y^2 - \frac{3x^2}{4} + (6 T_{1,0} - 21/2) y^3
 -  9y^4 (495 T_{1,0} - 220 T_{2,0} - 2296) / 220.
\label{eq6.8}
\end{equation}
One can similarly utilize Eqs.\ (\ref{eq4.16}), (\ref{eq4.17}) and (\ref{eq4.18}) for $C(y, T_{1,0}, T_{2,0})$, $D(y, T_{1,0}, T_{2,0})$, and $E(y, T_{1,0}, T_{2,0})$, respectively, in order to express the fourth derivative condition (\ref{eq2.6}) as
\begin{equation}
y  =  \frac{11}{3} B(y, T_{1,0}, T_{2,0}) + \frac{35}{3} C(y, T_{1,0}, T_{2,0})
 +  20 D(y, T_{1,0}, T_{2,0}) + 16 E(y, T_{1,0}, T_{2,0}).
\label{eq6.9}
\end{equation}
Given the prior determination of $T_{1,0} = 2.5533$, $K = -0.05793$, and given the physical couplant values $x(v) = 0.0253$, $z(v) = 0.0329$, Eqs.\ (\ref{eq6.8}) and (\ref{eq6.9}) represent two equations in the two unknowns $y$, $T_{2,0}$.  The smallest positive real solution for $y$ is $y = 0.05362$, in which case $T_{2,0} = -8.1744$ and, from Eq.\ (\ref{eq2.9}), $V^{\prime\prime} (v) = (219.5$ GeV)$^2$, results almost identical to the SFTP results of Section 4 [$T_{2,0} = - 8.1770$, $y_{N^2 LL} = 0.05392$, $V^{\prime\prime}(v) = \left(225\,{\rm GeV}\right)^2$].

One can continue this procedure through two subsequent orders of the $SFTP$.  The results one obtains are tabulated in Table \ref{vtab2}.  These results exhibit stability about $y = 0.054$, $[V^{\prime\prime}(v)]^{1/2} \cong 221$ GeV through four subleading orders in $y$.

\begin{table}[hbt]
\centering
\begin{tabular}{||c|c|c|c||}     \hline\hline
 $k$ & $y(v)$ & $T_{k,0}$ & $\left[ V^{\prime\prime}(v)\right]^{1/2}$ (GeV) \\ \hline\hline
0  & 0.05383 & 1 & 215.8 \\ \hline
1  & 0.05351 & 2.5533 & 221.7 \\ \hline
2 & 0.05362 & -8.1744 & 219.5   \\ \hline
3  & 0.05355 & 83.190 & 221.3 \\ \hline
4 & 0.05338 & -982.21  & 223.6\\ \hline\hline
\end{tabular}
\caption{Perturbative stability of results inclusive of $N^k LL$ contributions to the effective potential from the dominant couplant $y(v) \equiv \lambda(v) / 4\pi^2$ augmented by $LL$ contributions from the $t$-quark Yukawa couplant $x(v) \equiv
g_t^2 (v) / 4\pi^2$ and $z(v) = \alpha_s (v) / \pi$.}
\label{vtab2}
\end{table}

We conclude this section by developing a fully $NLL$ set of predictions in the couplants $\{x,y,z\}$.  To begin, we note that the additional terms involving $x$ and $z$ contributing to $B$ to $NLL$ order are
\begin{gather}
\Delta B_{NLL} = \Delta B_{SFTP} + \Delta_x B
\label{eq6.10}
\\
\Delta_x B  =  \left( - \frac{27}{4} + \frac{3 T_{1,0}}{2} \right) xy^2 + \frac{3}{2} x^3 - x^2 z + \frac{3}{4} xy^2
 +  U \left[ 3x^3 / 4 - 4x^2 z  \right] .
\label{eq6.11}
\end{gather}
In the above expression, one sees that $\Delta_x B \rightarrow 0$ as the $t$-quark Yukawa couplant $x$ goes to zero.

Similarly, we can write
\begin{gather}
\Delta C_{NLL} = \Delta C_{SFTP} + \Delta_x C,
\label{eq6.12}
\\
\Delta D_{NLL} = \Delta D_{SFTP} + \Delta_x D,
\label{eq6.13}
\\
\Delta E_{NLL} = \Delta E_{SFTP} + \Delta_x E.
\label{eq6.14}
\end{gather}
By subtracting from Eqs.\ (\ref{eq5.25}), (\ref{eq5.30}), and (\ref{eq5.34}) the explicitly $NLL$ SFTP contributions to Eqs.\ (\ref{eq4.9}), (\ref{eq4.10}) and (\ref{eq4.11}), the $NLL$ contributions to $C$, $D$ and $E$ arising entirely from the Yukawa sector are given by
\begin{gather}
\Delta_x C =   \Delta C_{NLL} - y^4 \left( 27 T_{1,0} - 621/8\right),
\label{eq6.15}
\\
\Delta_x D =   \Delta D_{NLL} - 9y^5 \left( -89 + 24 T_{1,0} \right) / 2 ,
\label{eq6.16}
\\
\Delta_x E = \Delta E_{NLL} - 27 y^6 \left( 240 T_{1,0} - 1049 \right) / 16.
\label{eq6.17}
\end{gather}
Now, if one uses only $NLL$ expressions for $\{B,C,D,E \}_{SFTP}$ in Eqs.\ (\ref{eq6.11})--(\ref{eq6.14}), one finds that

\begin{equation}
	\left( 
		\begin{array}{c}
			B\\C\\D\\E
		\end{array}
	\right)
=
	\left(
		\begin{array}{c}
			B_{LL}\\C_{LL}\\D_{LL}\\E_{LL}
		\end{array}
	\right)
+
\left(
		\begin{array}{c}
			\Delta B\\ \Delta C\\ \Delta D\\ \Delta E
		\end{array}
	\right)_{SFTP}
+
	\left(
		\begin{array}{c}
			\Delta_x B \\ \Delta_x C \\ \Delta_x D \\ \Delta_x E
		\end{array}
	\right)
=
	\left(
		\begin{array}{c}
			B_{LL} + \Delta B_{NLL} \\ C_{LL} + \Delta C_{NLL} \\
			D_{LL} + \Delta D_{NLL} \\ E_{LL} + \Delta E_{NLL}
		\end{array}
	\right) ,\\
\label{eq6.18}
\end{equation}
thereby recovering the series coefficients within Eq.\ (\ref{eq5.2}) for the potential $V_{NLL}$ (\ref{eq5.5}), as calculated in the remainder of Section 5.  

Applying the minimization condition (\ref{eq2.5}) to these series coefficients leads to an explicit expression for the unknown finite counterterm coefficient $U$ associated with the finite part chosen from Fig. 2:
\begin{equation}
\begin{split}
U  = & - \Biggl[ y^3 \left( 24 T_{1,0} - 42\right) + xy^2 \left( 6 T_{1,0} - 27 \right) + x^2 y \left( -6 T_{1,0} +3\right) + 6x^3 - 4x^2 z\Biggr.
\\
& \qquad+  \Biggl. 12y^2 - 3x^2 + 8 (y+K) \Biggr] / \left[x^2 (-16z + 3x )\right].
\end{split}
\label{eq6.19}
\end{equation}
We already know that $K = -0.057935$ from the $LL$ calculation of refs.\ \cite{5,6}.  We also have the $NLL$ results that $T_{1,0} = 2.5533$, as obtained prior to ``turning on'' $NLL$ Yukawa sector contributions $\Delta_x \{B,C,D,E\}$.   Given  physical values (\ref{eq2.1}) for $x$ and $z$, we see that Eq.\ (\ref{eq6.19}) is a relation expressing the finite counterterm coefficient $U \; (=D_{2,0,0,0})$ as a function of the only remaining unknown $y$. We substitute Eq.\ (\ref{eq6.19}) to eliminate $U$ from $\Delta B_{NLL}$ (\ref{eq5.21}), $\Delta C_{NLL}$ (\ref{eq5.26}), $\Delta D_{NLL}$ (\ref{eq5.31}) and $\Delta E_{NLL}$ (\ref{eq5.34}). We then find from substituting Eq.\ (\ref{eq6.18}) into the condition (\ref{eq2.6}) that $y$ is the solution of a degree-six polynomial equation.  The only viable solution to this equation ({\it i.e.,} the smallest real positive root) is $y(v) = 0.05311$, in which case we see from Eq.\ (\ref{eq6.19}) that $U = -17.306$.  Substituting these numerical values into $B_{LL}$ (\ref{eq5.20}), $\Delta B_{NLL}$ (\ref{eq5.21}), $C_{LL}$ (\ref{eq5.25}) and $\Delta C_{NLL}$ (\ref{eq5.26}), the running Higgs boson mass to $NLL$ order in the three dominant electroweak couplants $\{x,y,z \}$ is found from Eq.\ (\ref{eq2.9}) to be $\left[ V_{NLL}^{\prime\prime} (v) \right]^{1/2} = 227.8$ GeV.  This is to be compared with the $216$ GeV $LL$ result for these same three couplants;  similarly the $NLL$ value $y = 0.0531$ is a controllable departure from the $LL$ result $y = 0.0538$ \cite{5,6} discussed in Section 2. Thus, the fully $NLL$ extension of the $LL$ results for $V_{xyz}$ presented in refs.\ \cite{5,6} lead to a very modest decrease in $y(v)$ and a 5\% increase in the running Higgs boson mass, indicative of order-by-order stability of $V_{xyz}$.	  


\section{Discussion}

\subsection{Turning on the Electroweak Gauge Couplants}

\renewcommand{\theequation}{7.\arabic{equation}}
\setcounter{equation}{0}

As noted in Section 2, the leading-logarithm contributions of the electroweak gauge coupling constants $g_2 (v)$ and $g^\prime (v)$ to $\{B, C, D, {\rm and}\; E \}$ are listed in the Erratum to 
ref.\ \cite{6}, where they are denoted as $\Delta_{ew} B$, $\Delta_{ew} C$, $\Delta_{ew} D$ and $\Delta_{ew} E$.  The aggregate $\phi^4$ counterterm $K$ is altered as well, as indicated in Eq.\ (\ref{eq2.11}).  To incorporate these additional (algebraically lengthy) corrections into the potential of the previous section, we modify the $LL$ expression for $B$ to include electroweak gauge coupling constant contributions
\begin{gather}
B_{LL} = 3y^2 - \frac{3x^2}{4} + \Delta_{ew} B,
\label{eq7.1}
\\
\Delta_{ew} B = \frac{3}{64} s^2 + \frac{9}{64} r^2 + \frac{3}{32} rs.
\label{eq7.2}
\end{gather}
We find from condition (\ref{eq2.5}) for the $LL$ analysis inclusive of $\Delta_{ew} C$, $\Delta_{ew} D$ and $\Delta_{ew} E$ that $K = -0.058703$ \cite{6}.  One can then repeat the analysis of the previous section to include the $LL$ electroweak couplings and Yukawa couplings, in conjunction with the $N^k LL$ scalar field theory projection of the effective potential. The results of this analysis are listed in Table \ref{vtab3}.

\begin{table}[hbt]
\centering
\begin{tabular}{||c|c|c|c||}     \hline\hline
 $k$ & $y(v)$ & $T_{k,0}$ & $\left[ V^{\prime\prime}(v)\right]^{1/2}$ (GeV) \\ \hline\hline
0  & 0.05448 & 1 & 218.3 \\ \hline
1  & 0.05415 & 2.5603 & 224.4 \\ \hline
2 & 0.05426 & -8.1773 & 222.1   \\ \hline
3  & 0.05419 & 83.195 & 224.0 \\ \hline
4 & 0.05406 & -1191.8  & 225.5\\ \hline\hline
\end{tabular}
\caption{$y_{N^k LL} + \{x,z,r,s\}_{LL}$; {\it i.e.} results
for the $k^{th}$ subleading order of the SFTP augmented
by leading logarithm contributions from the Yukawa
couplant $x$, the QCD couplant $z$, and the $SU(2) \times U(1)$ gauge couplants $r$ and $s$.}
\label{vtab3}
\end{table}

As is evident from the table, both $y(v)$ and the running Higgs boson mass are very stable as higher-order contributions from $y$ alone are incorporated.  Moreover, the ${\cal{O}}(y^{k+1} \phi^4)$ finite counterterm coefficients $T_{k,0}$ are found to be virtually the same as these coefficients in Table \ref{vtab1}, for which the subleading $SM$ couplants $\{x,z,r,s\}$ are assumed to be zero.

The calculation to $NLL$ order in $\{x,y,z\}$ in the previous section can also be supplemented with $LL$ contributions from the gauge couplants $\{r,s\}$, since $r(v) = 0.0109$, $s(v) = 0.00324$ are substantially smaller than $x(v) = 0.0253$, $z(v) = 0.0329$, $y(v) \cong 0.054$.  Given the prior determinations of $K = -0.058703$ and $T_{1,0} = 2.5603$ [Table \ref{vtab3}], one finds that $U = -17.857$, only a small departure from the value $(-17.31)$ obtained in the last section, and that $y(v) = 0.05374$, $[V^{\prime\prime} (v)]^{1/2} = 230.7$ GeV.

\subsection{The Next Order Physical Higgs Mass}

Unlike the case of conventional spontaneously broken symmetry (CSB), the Lagrangian for radiatively broken symmetry (RSB) does not contain a primitive $\phi^2$ term.  This means that an approach involving the next-order calculation of momentum-independent contributions from the Higgs boson self-energy is not viable, as counterterm subtractions of infinity cannot occur.  Instead, the next-order Higgs boson inverse propagator mass term, as in ref.\ \cite{13}, must remain the next-order expression $V^{\prime\prime}_{eff} (v)$. The kinetic term for the inverse propagator can be worked out by analogy to the RG-analysis of the kinetic term extracted by Politzer \cite{14} for the massless gauge boson propagator.  This kinetic term may be written as
\begin{equation}
\Gamma (p,\mu) = \left[ 1 + {\cal{C}}(x,y,z,r,s) \log \left( \frac{p^2}{\mu^2} \right) \right]p^2
\label{eq7.3}
\end{equation}
where
\begin{equation}
\left[ \mu \frac{\partial}{\partial \mu} + \left( \beta_x \frac{\partial}{\partial x} + \ldots + \beta_s \frac{\partial}{\partial s} \right) 
  -   2 \gamma_\phi (x,y,z,r,s) \right] \Gamma(p,\mu) = 0
\label{eq7.4}
\end{equation}
Our sign for the anomalous dimension differs from Politzer's because our RG equation (\ref{eq2.3}) is consistent with a negative sign
\begin{equation}
\mu \frac{d\phi}{d \mu} = -\phi \gamma_\phi
\label{eq7.5}
\end{equation}
in the chain rule decomposition of $\mu \frac{d}{d\mu} F(x,y,z,r,s; \mu; \phi)$.  In ref.\ \cite{12}'s seminal calculation, Politzer calculated the analog coefficient to ${\cal{C}}$ [Eq.\ (\ref{eq7.3})] in the massless gauge boson propagator, in order to determine the corresponding anomalous dimension $\gamma_A$.  In our case $\gamma_\phi$ is known.  Lowest-order application of Eq.\ (\ref{eq7.4}) onto Eq.\ (\ref{eq7.3}) yields
\begin{equation}
{\cal{C}} (x,y,z,r,s) = -\gamma_\phi = -  \left(3x(\mu) / 4 - 9r (\mu) / 16  
 -   3s(\mu) / 16 + {\cal{O}}(y^2)\right).
\label{eq7.6}
\end{equation}
Thus the full next-order inverse propagator for the Higgs field, with $\mu$ chosen to be equal to the $vev$ $v$ is given by
\begin{equation}
\Gamma (p^2, v) = \left[ 1 - [3x(v) / 4 - 9r(v) / 16 - 3s (v) / 16] \log \left( \frac{p^2}{v^2} \right) \right] p^2
- V_{eff}^{\prime\prime} (v).
\label{eq7.7}
\end{equation}

The zero of Eq.\ (\ref{eq7.7}) corresponds to the physical Higgs boson mass $m_H$, the Higgs propagator pole: $\Gamma(m_H^2 , v) = 0$.\footnote{As has been borne out in other contexts \protect\cite{18} we assume  this propagator pole to be gauge independent.}
One finds this value to be reduced by only $0.2 - 0.3$ GeV from values of $V_{eff}^{\prime\prime} (v)$ between 220 and 231 GeV for the physically known values $x(v) = 0.0253$, $r(v) = 0.0109$, $s(v) = 0.00324$ of the running Yukawa and $SU(2) \times U(1)$ gauge couplants.  Thus the next-order physical Higgs boson mass is effectively the same thing as our next order determination of $V_{eff}^{\prime\prime} (v)$, largely because of the near proximity of $V_{eff}^{\prime\prime} (v)$ with the magnitude of the $vev$ itself ($v = 246.2$ GeV).

\subsection{Large Coupling Phenomenology}

One of the distinguishing features of radiative symmetry breaking (RSB) over conventional symmetry breaking (CSB) is the large value of the quartic scalar field couplant $y$ $(=\lambda / 4\pi^2)$ for the former case.  In CSB, a large value of $\lambda$ necessarily implies a large Higgs boson mass, as well as the Goldstone boson equivalence theorem \cite{15, 16}.  This theorem implies that processes in the large $\lambda$ limit, such as $H \rightarrow (W^+ W^-, ZZ)$ or $W^+ W^- \rightarrow (W^+ W^-, ZZ)$ are dominated by their equivalents within the Goldstone/Higgs sector, as defined by the complex scalar doublet
\begin{equation}
\phi =
\left(
\begin{array}{c}
 w^- \\ (\phi_3 - i z) / \sqrt{2}	
\end{array}
\right), \; \; \; \; 
	\phi^+ =
\left(
\begin{array}{c}
 w^+ \\ (\phi_3 + i z) / \sqrt{2}	
\end{array}
\right),
\label{eq7.8}
\end{equation}
where
\begin{equation}
\phi^+ \phi  =  \left( w^+ w^- + \left( \phi_3^2 + z^2 \right) / 2 \right)
 \equiv  \left( \phi_1^2 + \phi_2^2 + \phi_3^2 + \phi_4^2\right) / 2 = \Phi^2 / 2
\label{eq7.9}
\end{equation}
when expressed in terms of real scalar fields $\left( w^- = \left( \phi_1 - i \phi_2 \right) / \sqrt{2}, \; z = \phi_4\right)$.  We choose the convention that $\phi_H = \phi_3 - v$ is the physical Higgs field;  {\it i.e.,} that $\phi_3$ is the component of the Higgs doublet which acquires a non-zero vacuum expectation value through spontaneous or radiative symmetry breaking.

Curiously, we have found almost a complete overlap of Higgs/Goldstone sector predictions in CSB and RSB, despite an earlier erroneous claim on our part \cite{5}.

Let us begin with the CSB tree potential
\begin{equation}
V_{CSB}  =  - M^2 \phi^+ \phi + \lambda (\phi^+ \phi)^2
 =  -\frac{M^2}{2} \Phi^2 + \frac{\lambda}{4} \left( \Phi^2 \right)^2 .
\label{eq7.10}
\end{equation}
The minimum of this potential occurs at $v = \sqrt{M^2 / \lambda}$ where $m_H^2 = 2M^2$.  Consequently, we obtain the CSB value of $\lambda = m_H^2 / 2v^2$.  One finds from the Goldstone/Higgs sector (as would be applicable to the following processes were $\lambda$ large) that leading contributions to the following CSB amplitudes are given by:
{\allowdisplaybreaks
\begin{gather}
H \rightarrow ZZ: \; \; T \left( \phi_H \rightarrow zz \right) \equiv \left.\frac{\partial^3 V_{CSB}}{\partial \phi_H \partial z^2}\right|_{\phi_H = \phi_1 = \phi_2 = z = 0}  = 2\lambda v = m_H^2 / v
\label{eq7.11}
\\
H \rightarrow W^+ W^- : ~T\left( \phi_H \rightarrow w^+ w^-\right)
\equiv\left( \frac{\partial^3 V_{CSB}}{\partial \phi_H \partial\phi_1^2} 
+ \frac{\partial^3 V_{CSB}}{\partial \phi_H \partial \phi_2^2}\right)_{\phi_H = \phi_1 = \phi_2 = z = 0}
 =  4 \lambda v = 2m_H^2 / v
\label{eq7.12}
\\
W^+ W^- \rightarrow ZZ :~T\left( w^+ w^- \rightarrow zz\right)
 \equiv \left. \frac{\partial^4 V_{CSB}}{\partial w^+ \partial w^- \partial z^2} \right|_{w^\pm = z = \phi_H = 0} 
 =  2\lambda = m_H^2 / v^2
\label{eq7.13}
\\
W^+ W^- \rightarrow W^+ W^- :~T\left( w^+ w^- \rightarrow w^+ w^-\right) 
 \equiv \left. \frac{\partial^4 V_{CSB}}{(\partial w^+)^2(\partial w^-)^2}\right|_{w^\pm = z = \phi_H = 0}  
 =  4\lambda = 2m_H^2 / v^2
\label{eq7.14}
\\
W^+ W^- \rightarrow HH:~T\left( w^+ w^- \rightarrow \phi_H\phi_H\right)
\equiv \left. \frac{\partial^4 V_{CSB}}{\partial w^+\partial w^-\left(\partial\phi_H\right)^2}\right|_{w^\pm = z = \phi_H = 0}   
=2\lambda=m_H^2/v^2
\label{wwHH}
\\
HH \rightarrow HH: ~T\left( \phi_H \phi_H \rightarrow \phi_H \phi_H\right)
 \equiv \left. \frac{\partial^4 V_{CSB}}{(\partial\phi_H)^4}\right|_{w^\pm = z = \phi_H = 0}
 =  6 \lambda = 3m_H^2 / v^2 .
\label{eq7.15}
\end{gather}
}

For the case of RSB , the effective potential is
\begin{equation}
V_{RSB} = \pi^2 (\Phi^2)^2 (A + BL + CL^2 + DL^3 + EL^4 + \cdots)
\label{eq7.16}
\end{equation}
where
\begin{equation}
L = \log \left( \frac{2 \phi\phi^+}{v^2}\right) = \log\left( \frac{\Phi^2}{v^2}\right),
\label{eq7.17}
\end{equation}
and where minimization requires that $A = -B/2$.  In addition, we have from Eq.\ (\ref{eq2.9}) that to lowest order
\begin{equation}
m_H^2  =  \left.\frac{\partial^2 V_{RSB}}{\partial \phi_H^2} \right|_{\phi_1 = \phi_2 = \phi_H = \phi_4 = 0} 
 =  8\pi^2 v^2 (B+C)
\label{eq7.18}
\end{equation}
The corresponding lowest order Higgs/Goldstone amplitudes for the RSB case to those presented in Eqs.\ (\ref{eq7.11})--(\ref{eq7.15}) are
{\allowdisplaybreaks
\begin{gather}
H \rightarrow ZZ:~ T(\phi_H \rightarrow zz )
 \equiv  \left.\frac{\partial^3 V_{RSB}}{\partial\phi_H \partial z^2} \right|_{\phi_H = \phi_1 = \phi_2 = z = 0} 
 =  4\pi^2 v (2A + 3B + 2C)
 =  8\pi^2 v (B+C) = \frac{m_H^2}{v},
\label{eq7.19}
\\
H \rightarrow W^+ W^- :~ T(\phi_H \rightarrow w^+ w^-)
 \equiv \left. \frac{\partial^3 V_{RSB}}{\partial \phi_H \partial \phi_1^2} + \frac{\partial^3 V_{RSB}}{\partial\phi_H \partial\phi_2^2} \right|_{\phi_H = \phi_1 = \phi_2 = z = 0}
 =  16\pi^2v (B+C) = 2m_H^2 / v
\label{eq7.20}
\\
\begin{split}
W^+ W^- \rightarrow ZZ :~ T(w^+ w^- \rightarrow zz)
 &\equiv \left. \frac{\partial^4 V_{RSB}}{\partial w^+ \partial w^- \partial z^2} \right|_{w^\pm = z = \phi_H = 0}\\
 &=  8\pi^2 (A+3B/2 + C)
 =  8\pi^2 (B+C) = m_H^2 / v^2
 \end{split}
\label{eq7.21}
\\
\begin{split}
W^+ W^- \rightarrow W^+ W^- :~ T(w^+ w^- \rightarrow w^+ w^-)
 &\equiv \left. \frac{\partial^4 V_{RSB}}{(\partial w^+)^2(\partial w^-)^2} \right|_{w^\pm = z = \phi_H = 0}
 \\
 &=  8\pi^2 (2A + 3B + 2C)
 =  16\pi^2 (B+C) = 2m_H^2 / v^2
 \end{split}
\label{eq7.22}
\\
\begin{split}
W^+ W^- \rightarrow HH :~ T(w^+ w^- \rightarrow \phi_H\phi_H)
 &\equiv \left. \frac{\partial^4 V_{RSB}}{\partial w^+\partial w^-(\partial \phi_H)^2} \right|_{w^\pm = z = \phi_H = 0}
 \\
 &=\pi^2\left(8A+28B+56C+48D\right)
 \\
 &=24\pi^2\left(B+C\right)+\pi^2\left(32C+48D\right)
 =3m_H^2/v^2+\pi^2\left(32C+48D\right)
 \end{split}
 \label{wwHH2}
\\
\begin{split}
HH \rightarrow HH :~T(\phi_H \phi_H \rightarrow \phi_H \phi_H)
 &\equiv \left. \frac{\partial^4 V_{RSB}}{\partial \phi_H^4} \right|_{w^\pm = z = \phi_H = 0}
  =  24\pi^2 \left( A+ \frac{25}{6} B + \frac{35}{3} C + 20D + 16E\right)\\
  &=  24\pi^2 \left( \frac{11B}{3} + \frac{35C}{3} + 20D + 16E\right)
 =  24\pi^2 y = 6\lambda_{RSB}.
\end{split}
\label{eq7.23}
\end{gather}
}
The results (\ref{eq7.19})--(\ref{eq7.23}) all utilized $A = -B/2$ [Eq.\ (\ref{eq2.5})] as well as Eq.\ (\ref{eq7.18}) for the RSB Higgs mass.  
Note that RSB results from Eqs.\ (\ref{eq7.19})--(\ref{eq7.22}) are in complete agreement with corresponding leading order CSB results from Eqs.\ (\ref{eq7.11})--(\ref{eq7.14}).
The result (\ref{eq7.23}) makes additional use of Eq.\ (\ref{eq2.6}) for the quartic couplant $y$.  Since $\lambda_{RSB} \cong 5\lambda_{CSB}$, we see that lowest-order Higgs-Higgs scattering will be enhanced by a factor of 25 in the RSB scenario. 
The RSB amplitude $W^+W^-\to HH$ is enhanced by more than a factor of 3 from the CSB amplitude as well.
If one substitutes the leading-logarithm SFTP coefficients $B=3y^2$, $C=9y^3$, $D=27 y^4$ with 
$y=\lambda_{RSB}/4\pi^2=0.0541$ [see the discussion at the beginning of Section 3] into 
Eqs.\ (\ref{eq7.18}) and (\ref{wwHH2}), then $m_H=221\,{\rm GeV}$, and the final line of (\ref{wwHH2})
is numerically equal to $2.98$.  By contrast, the corresponding CSB amplitude is found from 
Eq.\ (\ref{wwHH}) when $m_H=221\,{\rm GeV}$ to have a numerical value of $0.807$.  Thus the RSB amplitude is seen to be enhanced by a factor of $2.98/0.807=3.7$ relative to the CSB amplitude \cite{startle}.

These leading-logarithm SFTP results are corroborated by the leading-logarithm effective potential {\it inclusive} of the subdominant Standard Model couplants $x(v)=g_t^2/4\pi^2$, $z(v)=\alpha_s/\pi$,
$r(v)=g_2^2/4\pi^2$, $s(v)=g'^2/4\pi^2$.  The RSB Higgs boson mass is now $218\,{\rm GeV}$ 
(Table \ref{vtab3}).  The coefficients $C$ and $D$ are respectively found to be $0.00152$ and $0.000256$, and the numerical value of the final line of Eq.\ (\ref{wwHH2}) is $2.96$.  For CSB with the same Higgs mass ($218\,{\rm GeV}$), the numerical value of the same amplitude (\ref{wwHH}) is $0.785$. 
Thus, the Higgs/Goldstone  sector scattering amplitude associated with $W_L^+W_L^-\to HH$ is enhanced in RSB over CSB by a factor of $2.96/0.785=3.8$ in close agreement with the leading logarithm SFTP.
Identical enhancements characterize the Higgs/Goldstone sector amplitudes $ZZ\to HH$.
However, as noted above, all other amplitudes considered are found to be {\it the same} to lowest order in CSB and RSB scenarios.  In particular, the Higgs widths $H \rightarrow ZZ$, $H \rightarrow W^+ W^-$ do not differ to lowest order in CSB and RSB scenarios \cite{startle}.

Suppose a Higgs boson of mass $220\,{\rm GeV}$ were found in near-future collider experiments, as predicted by radiatively-broken symmetry.  This mass in and of itself would not be a confirmation of the radiative mechanism, as the Higgs boson mass is not determined by CSB and could fortuitously have the same empirical value as in radiatively-broken electroweak symmetry breaking.  However, if the scattering processes $\sigma\left(W_L^+W_L^-\to HH\right)$, $\sigma\left(ZZ\to HH\right)$, $\sigma\left(HW\to HW\right)$ were found to be enhanced from CSB expectations by an order of magnitude, such an enhancement in conjunction with the $220\,{\rm GeV}$ Higgs boson mass would be a strong signal that electroweak symmetry is broken radiatively rather than conventionally.  

\subsection{Methodology}

The salient result presented in this paper for radiative electroweak symmetry breaking is the manifest controllability of corrections to the scalar-field self-coupling $y(v)$ $(=\lambda (v) / 4\pi^2)$ and the running Higgs boson mass $\left[ V^{\prime\prime} (v) \right]^{1/2}$ when contributions of sequentially subleading logarithms are incorporated into the effective potential's perturbative series. Such contributions are obtained from higher-than-one-loop terms in the renormalization-group equation. Indeed, it was noted by Coleman and Weinberg \cite{1} that the one-loop effective potential they obtained diagramatically could have also been obtained directly via the renormalization group (``Callan-Symanzik'') equation. This is demonstrated explicitly in Ref.\ \cite{5}.

However, the {\it full} set of leading logarithm contributions [as opposed to just the one-loop logarithm term] to the radiatively broken effective potential expressed in terms of the Standard Model's largest couplants $\{x,y,z\}$
can also be obtained from one-loop RG functions \cite{5}.  In Sections 5 and 6 of the present paper, particularly as summarized in Eq.\ (\ref{eq6.9}), these results are extended to next-to-leading logarithm order through use of the $\{x,y,z\}$-sensitive portions of the Standard Model's {\it two-loop} RG functions, an advance in its own right in the formulation of radiative electroweak symmetry breaking.

We are aware of the unconventional methodology of the paper in establishing the values of RG-inaccessible finite $\phi^4$ counterterms.  The successive approximations to the full effective potential series delineated by equations (\ref{eq5.4})--(\ref{eq5.6}) for the dominant Standard Model couplants $\{x,y,z\}$, or alternatively the successive approximations (\ref{eq4.1})--(\ref{eq4.4}) to the full SFTP potential, are imposed upon us because the aggregate $\phi^4$ coefficient $K$ is itself ${\cal{O}}(|y|)$ in magnitude by virtue of the minimization condition (\ref{eq2.5}).  A conventional perturbative approach, in which $K$ might be identified as simply a next order ${\cal{O}}(y^2)$ coefficient is simply not feasible. Coleman and Weinberg \cite{1} were able to escape this conundrum only by assuming (as appropriate to the time of their paper) that ``wrong sign'' Yukawa couplant contributions to the effective potential were small, enabling $y$ (or $\lambda$) to be an ${\cal{O}}(g_2^4)$ quantity.  Such is not the case for the true Standard Model, as pointed out in Section 1.  Thus the successive approximations in which $K$ is defined to be a {\it sum} of $\phi^4$ counterterms [Eqs.\ (\ref{eq5.3}) and (\ref{eq3.4})] appear to be the only field-theoretically consistent way to reconcile an ${\cal{O}}(|y|)$ magnitude for $K$.

Of course $K$ itself and its constituent $\phi^4$ coefficients in Eq.\ (\ref{eq5.3}) are necessarily determined via a sequential procedure of consistently applied renormalization conditions, for which we have chosen repeated applications of Eqs.\ (\ref{eq2.5}) and (\ref{eq2.6}).  On the face of things, this procedure does not appear to be equivalent to order-by-order perturbative subtraction schemes, such as $\overline{MS}$.  For such approaches to be viable, successive finite counterterms must be ``next order.''  Since $K$, the first of these, is itself ${\cal{O}}(|y|)$ in magnitude, {\it any} perturbative approach identifying $K$ with an ${\cal{O}}(y^2)$ coefficient is inappropriate.  By contrast, the approach delineated in Eqs.\ (\ref{eq5.4})--(\ref{eq5.6}) and Eqs.\ (\ref{eq4.1})--(\ref{eq4.4}) converges on the full effective potential series via summation of successively subleading-logarithm contributions generated by their lead terms, namely successively higher-order purely-$\phi^4$ terms within the potential [{\it e.g.} $T_{k,0} y^{k+1}$ is the $\phi^4$ coefficient that serves as the lead term of the summation $y^{k+1} S_k (yL)$ in Eq.\ (\ref{eq3.8})].  All such $\phi^4$ coefficients, however, are included in the aggregate coefficient $K$ common to {\it every} approximation [{\it e.g.} Eq.\ (\ref{eq3.4}) and Eqs.\ (\ref{eq4.1})--(\ref{eq4.4})].  

Since $K$ is indeed an ${\cal{O}}(|y|)$ quantity by virtue of the opposite-sign $t$-quark's contribution swamping the ${\cal{O}}(g_2^4)$ contribution to the first leading logarithm of the effective potential, such successive approximations appear to be the only way we have been able to find to formulate a ``large couplant'' version radiative symmetry breaking.  Indeed, the approach we have developed constrains the effective potential
\begin{itemize}
\item[1)]to maintain a minimum at the physical electroweak vacuum expectation value $v = G_F^{-1/2} 2^{-1/4} = 246.2$ GeV,
\item[2)]to satisfy a renormalization group equation [Eqs.\ (\ref{eq2.3}) and, for the SFTP case, Eq.\ (\ref{eq3.5})] whose perturbative $\beta$- and $\gamma$-functions are calculated via $\overline{MS}$,
\item[3)]to maintain a consistent definition (\ref{eq2.6}) of the scalar-field self-interaction couplant $y$ at the $\mu = v$ scale, and
\item[4)]to generate predictions for the scalar field self-coupling $\lambda$ (or $y = \lambda / 4\pi^2$) and the running Higgs boson mass $\left[ V^{\prime\prime} (v) \right]^{1/2}$ that are both reasonable in magnitude [$y = 0.05$ is sufficiently small for the RG functions (\ref{eq3.6}) and (\ref{eq3.7}) to decrease term-by-term in magnitude] and stable under subsequent subleading-logarithm corrections.
\end{itemize}
Point 2 above is of particular importance --- our approach is rendered consistent with $\overline{MS}$ {\it by construction}.

We reiterate that the discovery of a Higgs boson mass in the 220 GeV region, as indicated in Section 6, is not in itself a definitive confirmation of radiative electroweak symmetry breaking.  Such a discovery would clearly point to radiative symmetry breaking only if accompanied by evidence for an anomalously large scalar field self-interaction coupling constant --- {\it i.e.,} a $\lambda$ five to six times larger than the conventional symmetry breaking prediction $\lambda = M_H^2 / 2v^2$.  We have argued that such  enhancements would manifest themselves in Higgs-Higgs scattering and,  to a lesser extent, in 
scattering processes such as $W^+W^-\to HH$, $ZZ\to HH$.  However, no such enhancements are evident in
lowest order expressions for the Higgs width or in processes such as $WW \rightarrow ZZ, WW$ with Higgs/Goldstone sector analogs.

\section*{Acknowledgements}

We are grateful for discussions with V.A.\ Miransky and M.\ Sher, and for support from the Natural Sciences and Engineering Research Council of Canada.

\end{document}